\documentclass[12pt,a4paper]{article}
\usepackage{amssymb}
\usepackage{amsmath}
\usepackage{amsfonts}
\usepackage{geometry}
\usepackage{setspace}

\newtheorem{theorem}{Theorem}

\newtheorem{corollary}{Corollary}

\newtheorem{lemma}{Lemma}

\newtheorem{proposition}{Proposition}

\input{tcilatex}

\begin{document}

\title{Uniform Quasi ML based inference for the panel AR(1) model.}
\author{Hugo Kruiniger\thanks{%
Address: hugo.kruiniger@durham.ac.uk; Department of Economics, 1 Mill Hill
Lane, Durham DH1 3LB, England. I thank N. Peyerimhoff for helpful
discussions. } \\
Durham University}
\date{This version: 12 December 2025}
\maketitle

\vspace{6.2cm}

\bigskip

\bigskip

\bigskip

\noindent JEL\ classification: C12, C13, C23.\bigskip


\setcounter{page}{0} \thispagestyle{empty}

\newpage

\baselineskip=20pt

\renewcommand{\baselinestretch}{1.5}

\begin{center}
\textbf{Abstract}
\end{center}

\vspace{1cm}

Maximum Likelihood (ML) offers attractive alternatives to Generalized Method
of Moments (GMM)\ estimators for dynamic panel data models. However, to date
no\linebreak identification robust inference methods exist that can be used
in conjunction with the ML estimators for these models. In this paper we
propose ML based inference methods for panel AR(1) models with arbitrary
initial conditions and heteroskedasticity that are\linebreak robust to the
strength of identification. We show that (Quasi) Lagrange Multiplier (LM)
tests and confidence sets (CSs) that use the expected Hessian rather than
the observed Hessian of the log-likelihood function have correct asymptotic
size and coverage\linebreak probability in a uniform sense, respectively.
Such Quasi LM\ tests and CSs are also robust to misspecification of the
distribution of the data and to heterogeneity, including heteroskedasticity.
We derive the power envelope of a Fixed Effects version of such an LM test
for hypotheses involving the autoregressive parameter when the average
information matrix is estimated by a centered OPG\ estimator and the model
is only second-order identified, and show that it coincides with the maximal
attainable power curve in the worst case setting. We also study the
empirical size and power properties of these (Quasi) LM tests and find that
the hypothesis that the (Quasi) LM test has correct size cannot be rejected.

\setcounter{page}{0} \thispagestyle{empty}\newpage

\section{Introduction}

This paper proposes new inference methods for panel AR models with arbitrary
initial conditions and heteroskedasticity and possibly additional regressors
that are robust to the strength of identification. Specifically, we consider
several Maximum Likelihood based methods of constructing tests and
confidence sets (CSs) and show that (Quasi) LM tests and CSs that use the
expected Hessian rather than the observed Hessian of the log-likelihood have
correct asymptotic size in a uniform sense.

There exists a vast literature devoted to estimation of versions of the
panel AR(1) model or, more generally, dynamic panel data models. The
estimators in this literature can roughly be classified into two groups:
Instrumental Variables/Generalized Method of Moments (IV/GMM) type
estimators and Maximum Likelihood (ML) type estimators. Seminal
contributions to the first group include Anderson and Hsiao (1981, 1982),
Holtz-Eakin et al. (1988), Arellano and Bond (AB, 1991), Arellano and Bover
(ABov, 1995) and Ahn and Schmidt (AS, 1995), while seminal contributions to
the second group include Chamberlain (1980), Anderson and Hsiao (1981,
1982), MaCurdy (1982) and Hsiao et al. (2002). The last two papers
considered Fixed Effects (FE) ML\ estimators, that is, estimators that are
based on data in first differences and are consistent under minimal
assumptions, whereas the other two papers in the second group considered
(correlated) Random Effects (RE) ML estimators.\footnote{%
The RE approach assumes finite second moments of the data in levels while
the FE approach only assumes finite second moments of the data in
differences, cf. Kruiniger (2001, 2022).} \footnote{%
The Transformed MLE of Hsiao et al. (2002) and the FEMLE in Kruiniger (2001,
2013, 2022) are the same estimator and are the FE counterpart of the REMLE.}

Kruiniger (2001) has shown that when the data are i.i.d. and normal, then
the RE and FE MLEs of the autoregressive parameter $\rho $ in a panel AR(1)
model with homoskedastic errors and $\left\vert \rho \right\vert <1$ are
asymptotically equivalent to optimal AS-type GMM\ estimators,\linebreak
which in this case also exploit moment conditions that rely on time-series
homoskedasticity, and that when $\rho =1,$ the information matrix associated
with the FEMLE for the panel AR(1) model with homoskedastic errors is
singular. Kruiniger (2013) and Alvarez and Arellano (2022) have developed
RE\ and FE MLEs for the panel AR(1) model that allows for time-series
heteroskedasticity. These estimators of $\rho $ remain large $N$, fixed $T$
consistent when the data are non-normal or i.h.d., where $N$ and $T$ are the
dimensions of the panel. Kruiniger (2013) and Ahn and Thomas (2023)\ have
shown that when $\rho =1$ and the errors bar possibly the error of the last
period are homoskedastic over time, then RE and FE MLEs of $\rho $ that
allow for time-series heteroskedasticity are $N^{1/4}$-consistent and have
non-normal\ limiting distributions. The slower rate of convergence is
related to the fact that in this case the information matrix is singular so
that $\rho $ is only second-order locally identified, cf. Sargan (1983) and
Rotnitzky et al. (2000).

Monte Carlo studies in Hsiao et al. (2002), Kruiniger (2013) and Hayakawa
and Pesaran (2015) have found that the (Quasi) MLEs for the panel AR(1) and
ARX(1) models have very good finite sample properties but that when $\rho $
is close to one, Wald tests for hypotheses about $\rho $ are oversized and
asymptotic confidence intervals based on the QMLEs of $\rho $ are too narrow
because $\rho $ is weakly identified. However, to date no iden-\linebreak
tification robust inference methods have been proposed that are related to
these QMLEs.

Weak identification including second- rather than first-order local
identification affect the rate of convergence and the limiting distributions
of the estimators and pose a challenge to conducting inference both in the
case of GMM and the ML method. For reliable inference it is important that
tests and/or confidence sets (CSs) have correct asymptotic size in a uniform
sense. As Andrews et al. (2020) explain, pointwise asymptotics often provide
very poor approximations to the finite-sample size when the test statistic
of interest has a discontinuous pointwise asymptotic distribution. For this
reason Wald-type tests and CSs for (hypotheses/parameter vectors that
include) $\rho $ will generally not have correct asymptotic size in a
uniform sense. Also likelihood based (Quasi) LR-type tests and CSs for
(hypotheses/parameter vectors that include) $\rho $ will in many cases not
have correct (locally) uniform asymptotic size, cf. Rotnitzky et al. (2000)
and Bottai (2003). In this paper we will instead propose tests and CSs for
(hypotheses/parameter vectors that include) $\rho $ that are based on
(Quasi) LM test statistics that use the expected Hessian of a RE or FE
log-likelihood and show that they have correct uniform asymptotic size.

Our approach generalizes that of Bottai (2003), who considered CSs based on
similar LM test statistics in the context of identifiable one-dimensional
parametric models with a smooth likelihood function and information equal to
zero at a critical point, in at least two ways. Firstly, we consider
multi-parameter models and hypotheses. To show that our (Quasi) LM tests and
CSs have correct uniform asymptotic size, we make use of the fact that in
suitably reparametrized versions of the panel AR(1) model, the parameters
other than $\rho $ are still strongly identified when $\rho =1.$ Secondly,
by using Quasi LM test statistics, we also allow for misspecification of the
distribution of the data and heterogeneity.

Bottai (2003) explains for the one-parameter case why an LM\ test that is
based on the observed Hessian rather than the expected Hessian will not have
correct uniform asymptotic size when the parameter, say $\theta $, is
second-order identified at a critical point $\theta ^{\ast }$. In this case
the LM\ test statistic does not converge in distribution to a $\chi ^{2}(1)$
random variable under the sequence $\theta _{n}=\theta ^{\ast }-cn^{-1/4}$
as the sample size $n$ tends to infinity although it does converge to a $%
\chi ^{2}(1)$ random variable under sequences $\theta _{n}=\theta ^{\ast
}-cn^{-b}$ with $b>1/4.$ The same situation arises in the multi-parameter
case.

We will now discuss alternative, GMM-based methods. When the autoregressive
parameter $\rho $ is local to unity, then the AB GMM\ estimator has poor
finite sample properties and is inconsistent due to weak instruments, cf.
ABov, Blundell and Bond (BB, 1998) and Kruiniger (K, 2009). If in addition
the individual time series are covariance stationary or the number of
pre-sample realizations is large, then the System estimator of Abov and BB
may have a non-normal limiting distribution and its rate of convergence may
depend on the choice of the weight matrix, cf. K2009. Furthermore, when
mean-stationarity holds but the ratio of the variance of the individual
effects to the variance of the idiosyncratic errors is large, then the AB
GMM and the System estimator can also suffer from a weak instruments problem
and be biased, cf. Hayakawa (2007), Bun and Windmeijer (2010) and Kruiniger
(2001, 2022). Incidentally, Bun and Kleibergen (2022) and Kruiniger (2022)
have shown that provided that $T$ is not too small, $\rho $ is identified,
albeit possibly only second-order identified when $\rho =1$, by a set of
linear and quadratic AS-type moment conditions that only depend on a lack of
serial correlation assumption for the errors and on differenced data and
hence do not require mean-stationarity to hold.

GMM-based tests (and CSs) for $\rho $ that have correct uniform asymptotic
size when $\rho =1$ include (CSs based on) the Newey and West (1987) type
GMM\ LM\ test(-statistic)s that exploit ABov and SYS moment conditions,
respectively, see Madsen (2003) and K2009, and identification-robust
test(-statistics)s such as the GMM AR test of Stock and Wright (2000) and
the KLM and GMM-CLR tests of Kleibergen (2005) that exploit AB, ABov, SYS
and AS moments conditions, cf. Bun and Kleibergen (2014).\footnote{%
These GMM LM\ tests have correct uniform asymptotic size when $\rho =1$
because the scaled sample moment conditions they depend on (evaluated at $%
\rho =1$) and their first derivates are asymptotically independent. As a
result these GMM\ LM test statistics have the same limiting distribution as
they would have under strong identification, i.e., $N(0,1)$ (or $\chi
^{2}(1) $ when squared), see Kruiniger (2009). However, in general GMM LM
tests are not robust to lack of identification. For instance, if $\rho =1,$
the GMM LM test that only exploits the moment conditions of AB does not have
correct uniform asymptotic size and shouldn't be used,\thinspace
cf.\thinspace Bond and Windmeijer (2005).} Results in Newey and Windmeijer
(NW, 2009) imply that most of these identification robust test statistics
also have correct asymptotic size under many weak moment conditions
asymptotics with a restriction on the relative rate at which the number of
moment conditions and $N$ tend to infinity. The only possible exceptions are
the tests that exploit the non-linear AS moment conditions when NW's second
assumption, in particular global identification, does not hold. However,
this only happens when $T\geq 3$ and the variances of the errors change over
time at a constant rate, cf. Alvarez and Arellano (2022). Andrews and
Guggenberger (2017) have shown that under rather general conditions the KLM
and GMM-CLR tests for nonlinear moment condition models have correct
asymptotic size as $N\rightarrow \infty $ although they note that for
GMM-CLR tests this result depends in the multi-parameter case on how the
conditioning statistic, upon which the GMM-CLR test depends, is weighted.

Bun and Kleibergen (2014) have shown for $T\geq 4$ that in a worst case
scenario, where the variance of the initial observations goes to infinity,
the true value of $\rho $ is equal to one and the errors are homoskedastic
so that $\rho $ is only second-order locally identified, the power envelope
of the KLM test based on the AS\ moment conditions and a centered optimal
weight matrix, viz. the AS\ KLM\ test, coincides with the maximal attainable
power curve for testing $H_{0}:\rho =1-cN^{-1/4},$ and that in such a
scenario the AS GMM\ AR\ and the AS\ GMM LM tests have less power than the
AS KLM test when $T\geq 5$.\footnote{%
In the worst case scenario effectively only moment conditions based on
second moments of differences of the data are exploited, which do not
require mean-stationarity when $\left\vert \rho \right\vert <1$.}\ The
analysis of local power of these tests that is provided in Bun and
Kleibergen (2014) is different from that in Dovonon et al. (2020). The
latter assumes that the true value of $\rho $ equals $1-cN^{-1/4}$ and tests 
$H_{0}:\rho =1$, and finds in a simulation study that the AS\ GMM\ AR test
has better power properties than the AS\ KLM\ and AS\ GMM-CLR tests.

Andrews et al. (2019) list some limitations and weaknesses of inference
procedures that are based on identification robust test statistics. Firstly,
the KLM and GMM-CLR tests for subvectors of $\theta $ generally no longer
have correct asymptotic size when the nuisance parameters are not strongly
identified, although the GMM AR\ test for subvectors of $\theta $ still has
correct asymptotic size in this case when the errors are homoskedastic, cf.
Guggenberger et al. (2012, 2019) and Kleibergen (2021). Secondly, a large
number of inference procedures have been proposed for overidentified models
with nonhomoskedastic errors, some of which are based on analogs or
generalizations of the GMM-CLR test, but there is no consensus on what
procedures to use in practice, beyond the recommendation to use methods that
are efficient when the instruments are strong. Recently, Andrews (2018)
introduced a widely applicable approach to detecting weak identification and
constructing two-step confidence sets that controls coverage distortions
under weak identification. In cases where the model is well identified, his
method indicates this and reports nonrobust confidence sets with probability
tending to 1. Finally, it has been found that increases in the number of
moment conditions can lead to size distortions of the identification robust
tests in finite samples, cf. Kleibergen and Mavroeidis (2009), Kleibergen
(2019) and Bun et al. (2020). The GMM LM test is also susceptible to this
problem. In their Monte Carlo studies based on various panel AR(1) and panel
ARX(1) models with parameter values that correspond to different degrees of
identification strength both Hayakawa and Pesaran (2015) and Bun and
Poldermans (2015) find evidence for size distortions for the GMM\ LM, GMM
AR, KLM and GMM-CLR tests that exploit different sets of moment conditions
(e.g. AB, AS, SYS and subsets and collapsed versions of these sets) and that
the distortions get worse as $T$ or the number of moment conditions
increases. The size distortions become smaller when $N$ gets larger. Bun et
al. (2020) show that when the number of moment conditions is moderately
large, then the GMM AR\ test that is based on a weighting matrix that uses
centered moment conditions is oversized, whereas the uncentered version of
the GMM AR test is conservative. Bun et al. (2020) also propose a
degrees-of-freedom corrected centered GMM\ AR test that has good size
properties.

The paper proceeds as follows. In section 2 we present the panel AR(1)
model, the assumptions, the (Q)ML estimators and their asymptotic properties
including their local-to-unit root limiting distributions. In section 3 we
discuss the asymptotic size properties of various ML\ based tests including
Quasi LM tests that use the expected Hessian of the RE\ or the FE
log-likelihood function. We also derive the power envelope of the FE version
of such an QLM test for $H_{0}:\rho =1-cN^{-1/4}$ when the average
information matrix is estimated by a centered OPG\ estimator and the model
is only second-order identified, and the maximal attainable power curve for
testing $H_{0}:\rho =1-cN^{-1/4}$ in the worst case scenario. In section 4
we conduct a Monte Carlo study into the empirical size and power properties
of the QLM tests. Section 5 concludes and the appendix contains the proofs.$%
\vspace*{-0.22in}$

\section{The panel AR(1) model$\protect\vspace*{-0.1in}$}

Consider the panel AR(1) model with individual effects:$\vspace*{-0.12in}$

\begin{equation}
y_{i,t}=\rho y_{i,t-1}+\eta _{i}+\varepsilon _{i,t},\text{ where }\eta
_{i}=(1-\rho )\mu _{i},\vspace*{-0.18in}  \label{mdl}
\end{equation}%
for $i=1,...,N$ and $t=2,...,T.$ When deriving the asymptotics results, we
let $N$ become large while the number of observations per `individual', $T,$
remains fixed. The analysis below can be extended straightforwardly for
models that also include strictly exogenous covariates; we have omitted them
from the model to keep the presentation simple.

We assume that $-1<\rho \leq 1.$ Note that the model can be rewritten as $%
y_{i,t}-\mu _{i}=\rho (y_{i,t-1}-\mu _{i})+\varepsilon _{i,t}$ and that the $%
\eta _{i}$ disappear (the $\mu _{i}$ drop out) when $\rho =1$. The parame-
trization $\eta _{i}=(1-\rho )\mu _{i}$ prevents the individual effects from
turning into individual trends at $\rho =1$ and thereby avoids a
discontinuity in the data generating process at $\rho =1$.

The vectors of idiosyncratic errors $\varepsilon _{i}=(\varepsilon _{i,2}$ $%
...$ $\varepsilon _{i,T})^{\prime }$ are independently distributed across
individuals and satisfy the following Standard Assumptions, SA$k$, for $k=2$
or $k=4$:$\vspace{-0.12in}$ 
\begin{equation}
E(\varepsilon _{i,t})=0\text{ and }E\left\vert \varepsilon _{i,t}\right\vert
^{k+\varsigma }<\infty \text{ for }i=1,...,N\text{ and }t=2,...,T,
\label{b4}
\end{equation}%
where $\varsigma \geq 0$ is an arbitrarily small constant. In the sequel SA2
is denoted by SA.

The individual effects can often be treated as random effects. In this case
we make the following Random Effects Assumptions, REA$k$, for $k=2$ or $k=4$:%
$\vspace{-0.12in}$ 
\begin{eqnarray}
&&(y_{i,1}\text{ }\eta _{i})^{\prime },\text{ }i=1,...,N,\text{ are i.i.d.
with }\sigma _{y}^{2}=Var(y_{i,1}),\text{ }E\left\vert y_{i,1}\right\vert
^{k+\varsigma }<\infty ,\text{ and\qquad }  \label{b1} \\
&&E(\mu _{i})=0,\text{ }\sigma _{\mu }^{2}=E(\mu _{i}^{2}),\text{ }%
E\left\vert \mu _{i}\right\vert ^{k+\varsigma }<\infty \text{ and }E(\mu
_{i}y_{i,1})=\sigma _{\mu y}\text{ when }\left\vert \rho \right\vert <1.%
\vspace{-0.12in}  \notag
\end{eqnarray}%
In addition, we let $\sigma _{\eta }^{2}=E(\eta _{i}^{2})$ and $\sigma
_{\eta y}=E(\eta _{i}y_{i,1}).$ The i.i.d. assumption in (\ref{b1}) is only
made for presentational convenience. It can easily be relaxed. The
assumption $E\left\vert \mu _{i}\right\vert ^{k+\varsigma }<\infty $ (for $%
k=2$ or $k=4$) ensures that under covariance stationarity the means of the
data, i.e., $\eta _{i}/(1-\rho )=\mu _{i},$ $i=1,...,N$, are drawn from a
distribution with a finite variance rather than a variance that tends to
infinity when $\rho $ approaches one.

Unlike the RE ML\ estimators, the FE ML\ estimators only exploit data in
first differences. This reflects the fact that the FE approach entails
making minimal assumptions about the $\mu _{i}$ and the $y_{i,1}.$ In the FE
case we assume that $v_{i,1}\equiv y_{i,1}-\mu _{i},$ $i=1,...,N$, satisfy a
Fixed Effects Assumption, FEA$k$, for $k=2$ or $k=4$, cf. Kruiniger (2013):$%
\vspace{-0.12in}$ 
\begin{eqnarray}
&&v_{i,1},\text{ }i=1,...,N,\text{ are i.i.d. with }\sigma
_{v_{1}}^{2}=Var(v_{i,1})\text{ and }  \label{b2} \\
&&E\left\vert v_{i,1}\right\vert ^{k+\varsigma }<\infty \text{ when }%
\left\vert \rho \right\vert <1.  \notag
\end{eqnarray}%
The i.i.d. assumption in (\ref{b2}) is made for presentational convenience.
It can be relaxed.

Suppose that $y_{i,1}$ depends on $\mu _{i}$ in a linear fashion, i.e. $%
y_{i,1}=E(y_{i,1})+\alpha _{1}\mu _{i}+\varepsilon _{i,1}$ with $%
E(\varepsilon _{i,1})=0,$ and $\mu _{i}\perp \varepsilon _{i,1}$. In the
important case that $\alpha _{1}=1,$ we have $\mu _{i}\perp v_{i,1}$ and FEA$%
k$ does not impose any restrictions on $\mu _{i}$ and $y_{i,1}$ other than
those on $y_{i,1}-\mu _{i}$. However when $\alpha _{1}\neq 1,$ FEA$k$
implies restrictions on the $\mu _{i}$ themselves$.$

In the sequel REA2 and FEA2 are denoted by REA and FEA, respectively.

We add assumption B (Basic), which in the RE\ case amounts to\vspace{-0.1in} 
\begin{equation}
\varepsilon _{i,s}\perp \varepsilon _{i,t}\text{ for }i=1,...,N\text{ and }%
t\neq s.\vspace{-0.1in}  \label{b3}
\end{equation}%
and\vspace{-0.1in} 
\begin{equation}
\varepsilon _{i,t}\perp y_{i,1}\text{ and }\varepsilon _{i,t}\perp \eta _{i}%
\text{ for }i=1,...,N\text{ and }t=2,...,T,  \label{b5}
\end{equation}%
and in the FE case amounts to (\ref{b3}) and\vspace{-0.1in} 
\begin{equation}
\varepsilon _{i,t}\perp v_{i,1}\text{ for }i=1,...,N\text{ and }t=2,...,T%
\text{ when }\left\vert \rho \right\vert <1.  \label{b7}
\end{equation}%
We sometimes use an augmented version of assumption B, i.e., assumption B$%
^{\prime }$\ that in addi-\linebreak tion to (\ref{b3}) and (\ref{b5})
assumes that $\varepsilon _{i,l}\varepsilon _{i,s}\varepsilon _{i,t}\perp
y_{i,1}$ for all $l,s,t\in \{2,...,T\}$ and $i=1,...,N$.

In the paper we will also make reference to the assumption of
mean-stationarity:\vspace{-0.1in}%
\begin{equation}
E(y_{i,1}-\mu _{i})=0\text{\quad and\quad }E(\mu _{i}(y_{i,1}-\mu _{i}))=0.
\label{mstat}
\end{equation}

We can allow for heteroskedasticity of the $\varepsilon _{i,t}$ in the
time-series dimension:\vspace{-0.1in}%
\begin{equation}
E(\varepsilon _{i,t}^{2})=\lambda _{t}^{2}<\infty ,\text{ for }i=1,...,N%
\text{ and }t=2,...,T.\text{ }  \label{b8}
\end{equation}%
In some cases we make the stronger assumption of Time Series
Homoskedasticity (TSH):\vspace{-0.1in} 
\begin{equation}
E(\varepsilon _{i,t}^{2})=\sigma ^{2}<\infty ,\text{ for }i=1,...,N\text{
and }t=2,...,T\text{.}  \label{b6}
\end{equation}%
Let $\overline{\sigma }_{s}^{2}=N^{-1}\sum_{i=1}^{N}E(\varepsilon
_{i,s}^{2}) $, then we say that assumption TSH$_{t}^{\ast }$ holds if and
only if $\overline{\sigma }_{s}^{2}=\overline{\sigma }_{2}^{2},$ for $%
s=3,...,t.$ We will often make a `homoskedasticity' assumption that is
weaker than TSH, namely TSH$^{\ast }$, which is short for TSH$_{T}^{\ast }.$
Similarly we can use a weaker version of time-series heteroskedasticity,
namely $\overline{\sigma }_{t}^{2}=\lambda _{t}^{2}<\infty ,$ for $%
t=2,...,T. $\vspace{-0.14in}

\subsection{ML estimators}

Direct application of the Maximum Likelihood method to the nonstationary
panel AR(1) model with RE will generally yield an inconsistent estimator for 
$\rho $ due to correlation between the individual effects ($\eta _{i}$) and
the regressors ($y_{i,t-1},$ $t=2,...,T$). However, a consistent ML
estimator for $\rho $ can be obtained after reformulating the model.
Following Chamberlain (1980) we can decompose the $\eta _{i}$ into a term
that depends on the initial observation, $y_{i,1},$ and a term that does
not: \footnote{%
For the sake of a simple exposition we assume that $E(y_{i,1})=0$. A
situation where $E(y_{i,1})\neq 0$ can be handled by including an intercept
term in (\ref{re1}) and in (\ref{cm1}).\smallskip \medskip} \vspace{-0.14in}%
\begin{equation}
\eta _{i}=\pi (1-\rho )y_{i,1}+(1-\rho )v_{i},\text{ }i=1,...,N,\text{ }
\label{re1}
\end{equation}%
where $v_{i}$ is a new individual effect with $E(v_{i})=0$, $\pi (1-\rho )=$
plim$_{N\rightarrow \infty }\sum_{i=1}^{N}(\eta _{i}y_{i,1})/$ $%
\sum_{i=1}^{N}y_{i,1}^{2}$ and plim$_{N\rightarrow \infty
}N^{-1}\tsum\nolimits_{i=1}^{N}(y_{i,1}v_{i})=0.$

Let $y_{i}=(y_{i,2}$ $...$ $y_{i,T})^{\prime }$ and $y_{i,-1}=(y_{i,1}$ $...$
$y_{i,T-1})^{\prime }$ and let $\iota $ denote a vector of ones. Then using
the decomposition of the `correlated effects' $\eta _{i}$ given in (\ref{re1}%
), we can rewrite the panel AR(1) model with RE\ as\vspace{-0.14in}%
\begin{eqnarray}
&&y_{i}=\rho y_{i,-1}+\pi (1-\rho )y_{i,1}\iota +u_{i},\quad \text{where}
\label{cm1} \\
&&u_{i}=(1-\rho )v_{i}\iota +\varepsilon _{i}\text{\quad with\quad }%
E(\varepsilon _{i}\varepsilon _{i}^{\prime })=\Psi (\zeta )=diag(\lambda
_{2}^{2},\lambda _{3}^{2},...,\lambda _{T}^{2}).  \notag
\end{eqnarray}

Let plim$_{N\rightarrow \infty
}N^{-1}\tsum\nolimits_{i=1}^{N}v_{i}^{2}=\sigma _{v}^{2}$ and $\Phi =(1-\rho
)^{2}\sigma _{v}^{2}\iota \iota ^{\prime }+\Psi (\zeta )$. Then it is easily
verified that plim$_{N\rightarrow \infty
}N^{-1}\tsum\nolimits_{i=1}^{N}(y_{i,0}\iota ^{\prime }\Phi ^{-1}u_{i})=0$
and plim$_{N\rightarrow \infty
}N^{-1}\tsum\nolimits_{i=1}^{N}(y_{i,-1}^{\prime }\Phi ^{-1}u_{i})=0,$ cf.
similar results under homoskedasticity in Blundell and Bond (1998). After
imposing the assumption that the error components are i.i.d. and normal,
i.e., $u_{i}\sim i.i.d.$ $N(0,\Phi )$, application of the ML method to (\ref%
{cm1}) yields the RE ML\ estimator of $\rho $, $\pi $, $\sigma _{v}^{2}$ and 
$\zeta =(\lambda _{2}^{2}$ $\lambda _{3}^{2}$ $...$ $\lambda
_{T}^{2})^{\prime }$. This estimator will still be consistent when the $%
\varepsilon _{i,t}$ are heteroskedastic across both dimensions of the panel.$%
\,$

When calculating the REMLE it is convenient to use the reparameterization $%
\tilde{\pi}=\pi (1-\rho )$ and $\widetilde{\sigma }_{v}^{2}=(1-\rho
)^{2}\sigma _{v}^{2}.$ Let $\theta _{0}=(\rho ,\tilde{\pi},\widetilde{\sigma 
}_{v}^{2},\zeta ^{\prime })^{\prime }.$ Then the log-likelihood function for
the above model will be denoted by $l_{RE}(\theta
)=\tsum\nolimits_{i=1}^{N}l_{RE,i}(\theta )$ where $\theta =(r,\widetilde{p},%
\widetilde{s}_{v}^{2},z^{\prime })^{\prime }$ and $l_{RE,i}(\theta )$ is the
contribution to $l_{RE}(\theta )$ from `individual' $i$. The REMLE of $%
\theta _{0}$ will be denoted by $\widehat{\theta }_{RE}$ or simply by $%
\widehat{\theta }.$

Hsiao et al. (2002) has proposed the Transformed MLE for the panel AR(1)
model, which can be viewed as the FE counterpart of the REMLE, cf. Kruiniger
(2001). One can obtain the FEMLE for the model that allows for
heteroskedasticity by replacing $\mu _{i}$ in (\ref{mdl}) by $y_{i,1}+v_{i},$
and imposing that the $(v_{i}$ $\varepsilon _{i}^{\prime })$ are i.i.d. and
normal with $v_{i}\perp \varepsilon _{i}$ when $\left\vert \rho \right\vert
<1$.\footnote{%
For the sake of a simple exposition we assume that $E(v_{i})=0$. A situation
where $E(v_{i})\neq 0$ can be handled by including an intercept term in (\ref%
{cm2}).\smallskip \medskip} This amounts to imposing the restriction $\pi =1 
$ on the model in (\ref{cm1}) and leads to the following formulation of the
nonstationary panel AR(1) model with FE:\vspace{-0.14in}%
\begin{eqnarray}
&&y_{i}=\rho y_{i,-1}+(1-\rho )y_{i,1}\iota +u_{i}\text{,\quad where }
\label{cm2} \\
&&u_{i}=(1-\rho )v_{i}\iota +\varepsilon _{i}\text{\quad with\quad }%
E(\varepsilon _{i}\varepsilon _{i}^{\prime })=\sigma _{i}^{2}\Psi (\zeta ),%
\text{ }  \notag
\end{eqnarray}%
and where $v_{i}=-v_{i,1}$ satisfy assumption (\ref{b7}). After imposing $%
u_{i}\sim i.i.d.$ $N(0,\Phi )$ with $\Phi =\widetilde{\sigma }_{v}^{2}\iota
\iota ^{\prime }+\Psi (\zeta )$, application of the ML method to (\ref{cm2})
yields the FE MLE of $\rho $, $\widetilde{\sigma }_{v}^{2}$, $\zeta
=(\lambda _{2}^{2}$ $\lambda _{3}^{2}$ $...$ $\lambda _{T}^{2})^{\prime }$.
This estimator will also be consistent when the $\varepsilon _{i,t}$ are
heteroskedastic across both dimensions of the panel.$\,$\ The log-likelihood
function for the above model will be denoted by $l_{FE}(\theta )$ where in
this case $\theta =(r,\widetilde{s}_{v}^{2},z^{\prime })^{\prime }$. \vspace{%
-0.14in}

\subsection{QML estimation}

In the previous subsection we allowed for cross-sectional heteroskedasticity
but otherwise assumed homogeneity of the distributions of the (standardized)
idiosyncratic errors. However, such strong distributional assumptions with
respect to the errors are almost never satisfied by panel data. K2013 has
shown that when the data exhibit heterogeneity the ML method still yields
consistent RE and FE Quasi ML\ estimators for $\rho $ if SA, REA (or FEA)
and B hold, and $T\geq 3$ when $\left\vert \rho \right\vert <1$ and $T\geq 4$
when $\rho =1$ (cf. Theorem 2 in K2013). Under stronger conditions on the $%
\varepsilon _{i},$ $y_{i,1}$ and $\mu _{i}$ (or $v_{i,1}$), namely under
SA4, REA4 (or FEA4), B and an appropriate Lindeberg condition, K2013 has
also shown that the first-order fixed parameter asymptotic distributions of
the RE and FE QMLEs of $\theta _{0}$ are normal when either $\rho =1$ and
the average information matrix and the Expected Hessian of the
log-likelihood function are nonsingular or when $\left\vert \rho \right\vert
<1$, that is, $\sqrt{N}(\widehat{\theta }-\theta _{0})\overset{d}{%
\rightarrow }N(0,H(\theta _{0})^{-1}G(\theta _{0})H(\theta _{0})^{-1})$ in
these cases, where $H(\theta _{0})$ is the asymptotic Hessian of the
log-likelihood function and $G(\theta _{0})$ is the asymptotic information
matrix (cf. Theorem 3 in K2013).\vspace{-0.08in}

\subsubsection{Asymptotic properties of the QMLEs when $\protect\rho =1$ and 
$TSH^{\ast }$ holds}

K2001 has shown that if $\rho =1,$ the $\varepsilon _{i}$ are i.i.d. and
normal, and assumption TSH holds and has been imposed on the likelihood
function, then the Expected Hessian of $N^{-1}l_{FE}(\theta )$ (and thus
also the information matrix) is singular. Ahn and Thomas (2023) and Bond et
al. (2005) have established an analogous result for the Expected Hessian of $%
N^{-1}l_{RE}(\theta ).$ In these cases standard ML theory cannot be used to
derive the limiting distribution of the MLE. Applying the asymptotic theory
recently developed by Rotnitzky, Cox, Bottai and Robins (2000, henceforth
RCBR) for cases with an information matrix of rank one less than full, Ahn
and Thomas (2023) have derived the non-normal limiting distribution of the
REMLE. The theory of RCBR is based on using suitable Taylor expansions of
the possibly reparametrized log-likelihood and score functions around the
value of the parameter vector $\theta $\ at which the information matrix is
singular, viz. $\theta _{\ast }.$

Unsurprisingly, when $\rho =1$ and TSH$^{\ast }$ holds but TSH\ has not been
imposed, the Expected Hessians of $N^{-1}l_{RE}(\theta )$ and $%
N^{-1}l_{FE}(\theta )$ are singular as well. By extending the theory of RCBR
to allow for i.h.d. data, K2013 has generalized the result of Ahn and Thomas
(2023) as follows.

Application of the asymptotic results of RCBR or their extension for i.h.d.
data to the (Q)MLEs requires that one reparametrizes the RE and FE models
described in section 2.1 so that three conditions, viz. (B1)-(B3), which are
stated just above Theorem 1, are satisfied. They have been verified in K2013
and are similar to conditions given in RCBR. The reparametrizations are such
that the scores at $\theta _{\ast }$ for all the new parameters but one are
linearly independent, the QMLEs of these parameters converge at rate $%
O_{p}(N^{-1/2})$ and the score at $\theta _{\ast }$ for the remaining new
parameter, which is $\rho $, is equal to zero w.p.1.

In the case of the RE(Q)MLE we need to use the following new parametrization
(indicated by the subscript $n$): $\theta _{0,n}=(\rho _{n},\widetilde{%
\sigma }_{v,n}^{2},\zeta _{n}^{\prime },\tilde{\pi}_{n})^{\prime }$ where $%
\rho _{n}=\rho ,$ $\widetilde{\sigma }_{v,n}^{2}=\widetilde{\sigma }%
_{v}^{2}/\sigma _{2}^{2}-(1-\rho ),$ $\zeta _{n}=(\sigma _{2,n}^{2},\sigma
_{3,n}^{2},...,\sigma _{T,n}^{2})^{\prime }$ with $\sigma _{t,n}^{2}=\sigma
_{t}^{2}/\rho ,$ $t=2,...,T,$ so that $\zeta _{n}=\zeta /\rho $, and $\tilde{%
\pi}_{n}=\tilde{\pi}-(1-\rho ).$ Under this reparametrization the regression
equation becomes $y_{i}-y_{i,1}\iota =\rho _{n}(y_{i,-1}-y_{i,1}\iota )+%
\tilde{\pi}_{n}y_{i,1}\iota +u_{i}.$ Noting that we can express the elements
of $\theta $ as functions of the elements of $\theta _{n}=(r_{n},\widetilde{s%
}_{v,n}^{2},z_{n}^{\prime },\widetilde{p}_{n})^{\prime },$ viz. $\theta
(\theta _{n})=(r_{n},r_{n}s_{2,n}^{2}(\widetilde{s}%
_{v,n}^{2}+(1-r_{n})),r_{n}z_{n}^{\prime },\widetilde{p}_{n}+(1-r_{n}))^{%
\prime },$ the reparameterized log-likelihood function is given by $%
l_{n,RE}(\theta _{n})=\sum_{i=1}^{N}l_{n,RE,i}(\theta _{n})$ where $%
l_{n,RE,i}(\theta _{n})=l_{RE,i}(\theta (\theta _{n})).\,$To obtain results
for the FE(Q)MLE one leaves out the last elements of $\theta _{0},$ $\theta
, $ $\theta _{0,n}$ and $\theta _{n},$ i.e., the elements that correpond to $%
\widetilde{\pi },$ $\widetilde{p},$ $\widetilde{\pi }_{n}$\ and $\widetilde{p%
}_{n}$, respectively. Hence it suffices to focus on the RE(Q)MLE.

Note that if $\rho =1$ and TSH$^{\ast }$ holds, then $\theta _{0}=\theta
_{0,n}=(1,0,\sigma ^{2}\iota ^{\prime },0)^{\prime }$ for some $\sigma ^{2}$%
. Thus $\theta _{\ast }=(1,0,\sigma ^{2}\iota ^{\prime },0)^{\prime }$ for
some $\sigma ^{2}$. We may partition $\theta _{0}$ as $\theta _{0}=(\rho
,\delta ^{\prime })^{\prime }$, partition $\theta _{\ast }$ as $\theta
_{\ast }=(\rho _{\ast },\delta _{\ast }^{\prime })^{\prime }=(1,\delta
_{\ast }^{\prime })^{\prime }$, partition $\theta _{0,n}$ as $\theta
_{0,n}=(\rho _{n},\delta _{n}^{\prime })^{\prime }$ and partition $\theta
_{n}$ as $\theta _{n}=(r_{n},d_{n}^{\prime })^{\prime }.$ We will use $%
l_{i}, $ $l_{n}$ and $l_{n,i}$ as short for $l_{RE,i}(\theta ),$ $%
l_{n,RE}(\theta _{n})$ and $l_{n,RE,i}(\theta _{n}),$ respectively. We also
define:%
\begin{eqnarray*}
S_{i,1} &=&\frac{1}{2}\frac{\partial ^{2}l_{n,i}}{\partial r_{n}^{2}}%
|_{\theta _{\ast }},\text{\quad }S_{i,2}=\frac{\partial l_{n,i}}{\partial
d_{n}}|_{\theta _{\ast }},\text{\quad }\medskip
S_{i}=(S_{i,1},S_{i,2}^{\prime })^{\prime }, \\
\mathcal{I} &=&\mathcal{I}_{\theta _{\ast }\theta _{\ast }}=\medskip
\lim_{N\rightarrow \infty }N^{-1}\sum\nolimits_{i=1}^{N}E_{\theta _{\ast
}}(S_{i}S_{i}^{\prime }), \\
\mathcal{H}_{1,1} &=&\lim_{N\rightarrow \infty }\frac{2}{4!}N^{-1}E_{\theta
_{\ast }}\frac{\partial ^{4}l_{n}}{\partial r_{n}^{4}}|_{\theta _{\ast }},%
\text{\quad }\vspace*{0.35in}\mathcal{H}_{1,2}^{\prime }=\mathcal{H}%
_{2,1}=\lim_{N\rightarrow \infty }\frac{1}{2!}N^{-1}E_{\theta _{\ast }}\frac{%
\partial ^{3}l_{n}}{\partial r_{n}^{2}\partial d_{n}}|_{\theta _{\ast }}, \\
\mathcal{H}_{2,2} &=&\lim_{N\rightarrow \infty }\frac{2}{2!}N^{-1}E_{\theta
_{\ast }}\frac{\partial ^{2}l_{n}}{\partial d_{n}\partial d_{n}^{\prime }}%
|_{\theta _{\ast }},\text{\quad }\vspace*{0.35in}\mathcal{H}=\mathcal{H}%
_{\theta _{\ast }\theta _{\ast }}=\left[ 
\begin{array}{cc}
\mathcal{H}_{1,1} & \mathcal{H}_{1,2} \\ 
\mathcal{H}_{2,1} & \mathcal{H}_{2,2}%
\end{array}%
\right] , \\
\mathcal{H}^{-1} &=&[\mathcal{H}^{k,l}]\text{ with }\dim (\mathcal{H}%
^{1,1})=1\text{ }\vspace*{0.2in}\text{and }\dim (\mathcal{H}^{2,2})=\dim
(d_{n}), \\
C_{1,1} &=&-\lim_{N\rightarrow \infty }\frac{1}{5!}N^{-1}E_{\theta _{\ast }}%
\frac{\partial ^{5}l_{n}}{\partial r_{n}^{5}}|_{\theta _{\ast }},\text{\quad 
}C_{1,2}^{\prime }=C_{2,1}=\medskip -\lim_{N\rightarrow \infty }\frac{1}{%
2\times (3!)}N^{-1}E_{\theta _{\ast }}\frac{\partial ^{4}l_{n}}{\partial
r_{n}^{3}\partial d_{n}}|_{\theta _{\ast }},\quad \\
C_{2,2} &=&-\lim_{N\rightarrow \infty }\frac{1}{2!}N^{-1}E_{\theta _{\ast }}%
\frac{\partial ^{3}l_{n}}{\partial r_{n}\partial d_{n}\partial d_{n}^{\prime
}}|_{\theta _{\ast }},\text{\quad }C=\medskip \left[ 
\begin{array}{cc}
C_{1,1} & C_{1,2} \\ 
C_{2,1} & C_{2,2}%
\end{array}%
\right] , \\
U_{1,N} &=&(\frac{1}{3!}N^{-1/2}\frac{\partial ^{3}l_{n}}{\partial r_{n}^{3}}%
|_{\theta _{\ast }},N^{-1/2}\frac{\partial ^{2}l_{n}}{\partial r_{n}\partial
d_{n}^{\prime }}|_{\theta _{\ast }})^{\prime },\quad U_{2,N}=(\frac{1}{2!}%
N^{-1/2}\frac{\partial ^{2}l_{n}}{\partial r_{n}^{2}}|_{\theta _{\ast
}},N^{-1/2}\frac{\partial l_{n}}{\partial d_{n}^{\prime }}|_{\theta _{\ast
}})^{\prime }, \\
U_{N} &=&U_{1,N}+C\mathcal{H}^{-1}U_{2,N},\quad U\sim N(0,\Sigma _{U}), \\
\breve{Z}_{N} &=&(\breve{Z}_{1,N},\breve{Z}_{2,N}^{\prime })^{\prime }=-%
\mathcal{H}^{-1}U_{2,N}\text{ with }\dim (\breve{Z}_{1,N})=1, \\
\breve{Z} &=&(\breve{Z}_{1},\breve{Z}_{2}^{\prime })^{\prime }\sim N(0,%
\mathcal{H}^{-1}\mathcal{IH}^{-1})\text{ with }\dim (\breve{Z}_{1})=1,
\end{eqnarray*}%
\begin{equation*}
\breve{B}\sim Bernoulli\text{ with }\Pr (\breve{B}=0)=\Pr (U^{\prime }\breve{%
Z}>0).
\end{equation*}%
Note that $U_{2,N}=(N^{-1/2}\tsum\nolimits_{i=1}^{N}S_{i,1},$ $%
N^{-1/2}\tsum\nolimits_{i=1}^{N}S_{i,2}^{\prime })^{\prime }$ and that
condition (B2), which requires that for all $K\in 
\mathbb{R}
^{T+1}$ $\frac{\partial ^{2}l_{n}}{\partial \rho _{n}^{2}}|_{\theta _{\ast
}}\neq K^{\prime }\frac{\partial l_{n}}{\partial \delta _{n}}|_{\theta
_{\ast }}$ with positive probability, implies that $\mathcal{I}$ is
nonsingular.

Condition (B3), which requires that for all $\overline{K}=(k$ $K^{\prime
})^{\prime }\in 
\mathbb{R}
^{T+2}$\ $\frac{\partial ^{3}l_{n}}{\partial \rho _{n}^{3}}|_{\theta _{\ast
}}\neq k\frac{\partial ^{2}l_{n}}{\partial \rho _{n}^{2}}|_{\theta _{\ast
}}+K^{\prime }\frac{\partial l_{n}}{\partial \delta _{n}}|_{\theta _{\ast }}$
with positive probability, ensures that $U_{N}$ is not identically equal to
zero.

Condition (B1) requires that $\frac{\partial l_{n}}{\partial \rho _{n}}%
|_{\theta _{\ast }}=0$ w.p.1.

In order to obtain the parametrization for which conditions (B1)-(B3) are
met\linebreak K2013 has used the formula $\theta _{n}=\theta +[0,\widetilde{K%
}^{\prime }]^{\prime }(r-1),$ where $\widetilde{K}=\tsum_{i}E((\frac{%
\partial l_{i}}{\partial \rho }|_{\theta _{\ast }})(\frac{\partial l_{i}}{%
\partial \delta }|_{\theta _{\ast }}))\times $\linebreak\ $[\tsum_{i}E((%
\frac{\partial l_{i}}{\partial \delta }|_{\theta _{\ast }})(\frac{\partial
l_{i}}{\partial \delta }|_{\theta _{\ast }})^{\prime })]^{-1}$, cf. RCBR p.
264.

Let $\mathbf{1}\{\breve{Z}_{1}>0\}=1$ if $\breve{Z}_{1}>0,$ let $\mathbf{1}\{%
\breve{Z}_{1}>0\}=0$ if $\breve{Z}_{1}\leq 0,$ let $\mathbf{1}\{\breve{Z}%
_{1}\leq 0\}=1-\mathbf{1}\{\breve{Z}_{1}>0\},$ and let $\theta _{a,n}=(\rho
_{n},\widetilde{\sigma }_{v,n}^{2},\zeta _{n}^{\prime })^{\prime }.$ Then
K2013 obtained the following result when the $\varepsilon _{i}$ (or,
equivalently, the data) are i.h.d.:\pagebreak

\begin{theorem}[K2013]
Let assumptions SA4, REA4 (FEA4) and B hold and let $T\geq 4$. Let $\rho =1$
and TSH$^{\ast }$ hold, that is, let $\theta _{0,n}=\theta _{\ast }$ for
some $\sigma ^{2}.$ Finally, let $\widehat{\theta }_{n}\equiv (\widehat{\rho 
}_{n},\widehat{\delta }_{n})^{\prime }$ be the REQMLE (FEQMLE) that
maximizes the value of $l_{n}(\theta _{n})=l_{n,RE}(\theta _{n})$ $%
(l_{n,FE}(\theta _{n}))$. If the regularity conditions (A1)-(A7) given in
K2013 hold and $\mathcal{H}$ is nonsingular, then \vspace{0.08in}\newline
a) $\frac{\partial l_{n}}{\partial r_{n}}|_{\theta _{\ast }}=0$ and $\frac{%
\partial l_{n}}{\partial d_{n}}|_{\theta _{\ast }}\neq 0$ a.s. Furthermore, $%
rank(I_{\delta _{n}\delta _{n}})=\dim (d_{n})$ a.s. \vspace{0.14in}\newline
b) $\left[ 
\begin{array}{c}
N^{1/4}(\widehat{\rho }_{n}-1) \\ 
N^{1/2}(\widehat{\delta }_{n}-\delta _{n})%
\end{array}%
\right] \overset{d}{\rightarrow }\left[ 
\begin{array}{c}
(-1)^{\breve{B}}\breve{Z}_{1}^{1/2} \\ 
\breve{Z}_{2}%
\end{array}%
\right] \mathbf{1}\{\breve{Z}_{1}>0\}+\left[ 
\begin{array}{c}
0 \\ 
\breve{Z}_{2}-(\mathcal{H}^{2,1}/\mathcal{H}^{1,1})\breve{Z}_{1}%
\end{array}%
\right] \mathbf{1}\{\breve{Z}_{1}\leq 0\}$\vspace{0.14in}\newline
$U_{N}\overset{d}{\rightarrow }U$ and $\breve{Z}_{N}\overset{d}{\rightarrow }%
\breve{Z},$ and if in addition assumption B$^{\prime }$ holds, then$%
\smallskip $\newline
c1) $\mathcal{I}_{RE}=diag(\mathcal{I}_{RE,\theta _{a,n}\theta _{a,n}},%
\mathcal{I}_{RE,\tilde{\pi}_{n}\tilde{\pi}_{n}})$ where $\dim (\mathcal{I}%
_{RE,\tilde{\pi}_{n}\tilde{\pi}_{n}})=1;$ $\mathcal{I}_{RE,\theta
_{a,n}\theta _{a,n}}=\mathcal{I}_{FE}.\smallskip $\newline
c2) $\mathcal{H}_{RE}=diag(\mathcal{H}_{RE,\theta _{a,n}\theta _{a,n}},%
\mathcal{H}_{RE,\tilde{\pi}_{n}\tilde{\pi}_{n}})$ where $\dim (\mathcal{H}%
_{RE,\tilde{\pi}_{n}\tilde{\pi}_{n}})=1;$ $\mathcal{H}_{RE,\theta
_{a,n}\theta _{a,n}}=\mathcal{H}_{FE}.\smallskip \newline
$d) in general $\widehat{\rho }_{RE}\overset{asy}{\nsim }\widehat{\rho }%
_{FE} $ but if the $\varepsilon _{i}$ are i.i.d. and normal, then $\widehat{%
\rho }_{RE}\overset{asy}{\sim }\widehat{\rho }_{FE}.$
\end{theorem}

The QMLEs of $\theta $ are given by $\widehat{\theta }=\theta (\widehat{%
\theta }_{n})$. The parameter $\rho $ is second-order identified (cf.
Sargan, 1983) because $\frac{\partial l_{n}}{\partial r_{n}}|_{\theta _{\ast
}}=0,$ while $\frac{\partial ^{2}l_{n,i}}{\partial r_{n}^{2}}|_{\theta
_{\ast }}\neq 0$, and because for all $K\in 
\mathbb{R}
^{T+1}$ $\frac{\partial ^{2}l_{n}}{\partial r_{n}^{2}}|_{\theta _{\ast
}}\neq K^{\prime }\frac{\partial l_{n}}{\partial d_{n}}|_{\theta _{\ast }}$
with positive probability, that is, (B.2) holds. As a result the expansion
of the log-likelihood involves $N^{1/4}(\widetilde{\rho }-1)$ rather than $%
N^{1/2}(\widetilde{\rho }-1)$ and the QMLEs of $\left\vert \rho
-1\right\vert $ converge at rate $O_{p}(N^{-1/4}).$ Furthermore, when $%
\theta _{0,n}=\theta _{\ast }$ and $\breve{Z}_{1,N}>0,$ $l_{n}(\theta _{n})$
is bimodal. In this case the QMLE of $\rho $ is determined by higher-order
terms in the expansion of $l_{n}(\theta _{n})$ around $\theta _{n}=\theta
_{\ast }$, i.e., $U_{N}^{\prime }\breve{Z}_{N}$, cf. K2013. Although $\frac{%
\partial l_{n}}{\partial d_{n}}|_{\theta _{\ast }}\neq 0$ a.s.\ and $%
\widehat{\delta }_{n}-\delta _{n}=O_{p}(N^{-1/2}),$ whereas $\frac{\partial
l_{n}}{\partial r_{n}}|_{\theta _{\ast }}=0$ a.s.\ and $\widehat{\rho }%
_{n}-1=O_{p}(N^{-1/4}),$ all elements of $\widehat{\theta }_{a,n}$ have a
non-normal limiting distribution. All these findings would still hold if the
model also included exogenous regressors.

Theorem 1 holds for i.h.d. data. In the special case where the $\varepsilon
_{i}$ are i.i.d. and normal, $\mathcal{H}=-\mathcal{I}$, $\breve{Z}\sim N(0,%
\mathcal{I}^{-1}),$ $U\perp \breve{Z}$ and hence $\Pr (\breve{B}=0|\breve{Z}%
_{1})=1/2,$ $\widehat{\rho }$ has a symmetric limiting distribution, and $%
\widehat{\rho }_{RE}$ is asymptotically equivalent to $\widehat{\rho }_{FE}$%
. However, in general (i.e., except for some special cases) if the $%
\varepsilon _{i}$ are i.i.d and non-normal or i.h.d., then $\mathcal{H}\neq -%
\mathcal{I}$, $E(U\breve{Z}^{\prime })\neq 0$ and hence $\Pr (\breve{B}=0|%
\breve{Z}_{1})\neq \Pr (\breve{B}=0)$, and $\widehat{\rho }$ has an
asymmetric limiting distribution. Moreover, $\widehat{\rho }_{RE}$ is not
asymptotically equivalent to $\widehat{\rho }_{FE}$.

\subsection{The distributional properties of QML estimators when $\protect%
\rho $ is close to unity}

It is well-known that (Quasi) ML estimators can be reinterpreted as GMM
estimators, cf. e.g. Newey and McFadden (NMcF, 1994). The underlying moment
conditions can be obtained by setting the expected score vector equal to
zero. It follows that the Expected Hessian of the (quasi) log-likelihood
function equals the first derivative of the vector of moment conditions
exploited by the (Q)MLE with respect to the parameters. Therefore when the
Expected Hessian of the (quasi) log-likelihood function is almost singular,
the (Q)MLE suffers from a `weak moment conditions problem.' K2013 gives
necessary and sufficient conditions for this situation to arise when $%
Var(y_{i,1}-\mu _{i})\propto (1-\rho )^{0}=1$ $\forall i\in \{1,2,...,N\}$:

\begin{theorem}[K2013]
Let assumptions SA, REA (or FEA) and B hold and let\linebreak $%
Var(y_{i,1}-\mu _{i})\propto (1-\rho )^{0}=1$ $\forall i\in \{1,2,...,N\}$.
Furthermore let $\rho $ be local to unity. Then the Expected Hessian of $%
l_{RE}(\theta )$ $(l_{FE}(\theta ))$ is almost singular if and only if
either assumption TSH$^{\ast }$ holds, assumption TSH has been imposed on $%
l_{RE}(\theta )$ $(l_{FE}(\theta ))$ and $T\geq 3,$ or assumption TSH$%
_{T-1}^{\ast }$ (almost) holds and $T\geq 4$.
\end{theorem}

If the Expected Hessian of the (quasi) log-likelihood function is almost
singular, one can obtain a better approximation to the finite sample
distribution of the (Q)MLE than the usual approximation by using local
asymptotics. K2013 obtained the following result:

\begin{theorem}[K2013]
Let assumptions SA4, FEA4 (or REA4) and B hold, let $T\geq 4$, let $\delta
_{\ast }$ be such that $(1,\delta _{\ast }^{\prime })^{\prime }=\theta
_{\ast },$ and let $\{\theta _{n,N}\}$ be such that $N^{1/4}(\rho
_{n,N}-1)=o(1)$ and $N^{1/2}(\delta _{n,N}-\delta _{\ast })=o(1)$ for some $%
\sigma ^{2}.$ Let $\{\theta _{i,n,N}\},$ $i=1,...,N,$ be such that $\rho
_{i,n,N}=\rho _{n,N},$ $N^{-1}\sum_{i=1}^{N}\widetilde{\sigma }%
_{v,i,n,N}^{2}=\widetilde{\sigma }_{v,n,N}^{2},$ $N^{-1}\sum_{i=1}^{N}\zeta
_{i,n,N}=\zeta _{n,N}$ $($and $\tilde{\pi}_{i,n,N}=\tilde{\pi}_{n,N}).$ Let $%
\widehat{\theta }_{n}=(\widehat{\rho }_{n},\widehat{\delta }_{n})^{\prime }$
be the FEQMLE (REQMLE) that maximizes the value of $l_{n}(\theta
_{n})=l_{n,FE}(\theta _{n})$\linebreak $(l_{n,RE}(\theta _{n}))$. Let $%
\mathcal{H},$ $\breve{B}$ and $\breve{Z}=(\breve{Z}_{1},\breve{Z}%
_{2}^{\prime })^{\prime }$ satisfy the definitions given just above theorem
1. If $\mathcal{H}$ is nonsingular and $(y_{i,1}$ $y_{i}^{\prime })^{\prime
} $ is generated under $\theta _{i,n,N},$ $i=1,...,N,$ then\vspace{0.06in}%
\newline
a) $\widehat{\theta }_{n}-\theta _{\ast }=o_{p}(1).$\vspace{0.06in}\newline
b) $\left[ 
\begin{array}{c}
N^{1/4}(\widehat{\rho }_{n}-1) \\ 
N^{1/2}(\widehat{\delta }_{n}-\delta _{\ast })%
\end{array}%
\right] \overset{d}{\rightarrow }\left[ 
\begin{array}{c}
(-1)^{\breve{B}}\breve{Z}_{1}^{1/2} \\ 
\breve{Z}_{2}%
\end{array}%
\right] \mathbf{1}\{\breve{Z}_{1}>0\}+\left[ 
\begin{array}{c}
0 \\ 
\breve{Z}_{2}-(\mathcal{H}^{2,1}/\mathcal{H}^{1,1})\breve{Z}_{1}%
\end{array}%
\right] \mathbf{1}\{\breve{Z}_{1}\leq 0\}$\vspace{0.14in}\newline
c) in general $\widehat{\rho }_{RE}\overset{asy}{\nsim }\widehat{\rho }_{FE}$
but if assumption B$^{\prime }$ also holds and the $u_{i}$ are i.i.d. and
normal, then $\widehat{\rho }_{RE}\overset{asy}{\sim }\widehat{\rho }_{FE}.$
\end{theorem}

Note that since $\widetilde{\sigma }_{v,n}^{2}=\widetilde{\sigma }%
_{v}^{2}/\sigma _{2}^{2}-(1-\rho )=(1-\rho )((1-\rho )\sigma _{v}^{2}/\sigma
_{2}^{2}-1)$ and $\tilde{\pi}_{n}=(\pi -1)(1-\rho )$, both condition $%
N^{1/2}(\widetilde{\sigma }_{v,n,N}^{2}-0)=o(1)$ and condition $N^{1/2}(%
\tilde{\pi}_{n,N}-0)=o(1)$ normally require that $N^{1/2}(\rho
_{n,N}-1)=o(1) $ rather than $N^{1/4}(\rho _{n,N}-1)=o(1).$ However, when
for instance $\pi =1$ and $\sigma _{v}^{2}=2\sigma _{2}^{2}/(1-\rho ^{2}),$
then the results of theorem 3 hold under any $\{\rho _{n,N}\}$ such that $%
N^{1/4}(\rho _{n,N}-1)=o(1).$ In any case, when both $N^{1/4}(\rho
_{n,N}-1)=o(1)$ and $N^{1/2}(\delta _{n,N}-\delta _{\ast })=o(1)$ for some $%
\sigma _{2}^{2},$ the RE-\ and the FEQMLE of $\rho $ are $N^{1/4}-$%
consistent.

Results in RCBR suggest that in principle under $(\rho _{n,N}-1)=-\lambda
N^{-b}$ different local asymptotic distributions and rates of convergence
are obtained for the QMLE of $\rho $ when $b=1/4,$ $1/6<b<1/4$, $b=1/6$ or $%
0<b<1/6,$ respectively.

\section{Likelihood based tests}

Wald test statistics, some versions of (Quasi) LM test statistics, and
(Quasi) LR test statistics that are used for testing an hypothesis about $%
\rho $ and are based on the reparametrized RE or FE\ likelihood do not
uniformly converge to their fixed parameter first-order limiting
distributions near the singularity point, cf. Bottai (2003). As a
consequence these tests do not have correct asymptotic size in a uniform
sense. (Q)LM test statistics that are standardised by (using a sandwich
formula involving) the \emph{expected} rather than the observed average
Hessian are exceptions to this rule. Bottai (2003) explains why this is the
case in the context of models with a single parameter. As we will show in
the proof of theorem 4, when testing a hypothesis that includes a
restriction on the parameter that has zero score at the singularity point,
these test statistics still converge uniformly to their fixed parameter
first-order limiting distribution, which is a central $\chi ^{2}$%
-distribution, near the singularity point and these tests have correct
asymptotic size in a uniform sense. Crucially, the expected average Hessian
of $l_{n}(\theta _{n})$, i.e., $\overline{H}(\mathcal{\breve{\theta}}%
_{n})\equiv E_{\mathcal{\breve{\theta}}_{n}}(\frac{1}{N}\frac{\partial
^{2}l_{n}(\theta _{n})}{\partial \theta _{n}\partial \theta _{n}^{\prime }}%
|_{\mathcal{\breve{\theta}}_{n}})$, is\linebreak negative definite and hence
nonsingular for any value of $\mathcal{\breve{\theta}}_{n}$ that differs
from the singu- larity point $\theta _{\ast }$. Note that the values of the
elements of $\overline{H}(\mathcal{\breve{\theta}}_{n})$ do not depend on
the true distribution of the data. Alternative testing approaches that are
based on a QLR test statistic but use a so-called type 2 robust critical
value, which is determined by a statistic that indicates the closeness to
the singularity point, or a least favourable critical value (cf. Andrews and
Cheng, 2013) either do not work due to non-monotonicity of the limiting
distributions of the QLR test statistic near the singularity point or are
conservative and have low power.

We will now introduce the QLM test statistic $QLM(\widetilde{\mathcal{\theta 
}}_{n})$ for testing $H_{0}:$ $A\mathcal{\theta }_{0,n}=a$, which includes a
restriction on $\rho $, where $\widetilde{\mathcal{\theta }}_{n}$ is a
restricted QML\ estimate of $\mathcal{\theta }_{0,n}$ such that $A\widetilde{%
\mathcal{\theta }}_{n}=a,$ $A$ is a $J\times \dim (\mathcal{\theta })$
constant matrix of rank $J,$ and $a\ $is a constant vector. Define the
average information matrix $\overline{\mathcal{J}}\mathcal{(\mathcal{\breve{%
\theta}}}_{n}\mathcal{)}=N^{-1}\tsum_{i=1}^{N}\mathcal{J}_{i}\mathcal{(%
\mathcal{\breve{\theta}}}_{n}\mathcal{)}$ with $\mathcal{J}_{i}\mathcal{(%
\breve{\theta}}_{n}\mathcal{)}=\left( \frac{\partial l_{n,i}(\theta _{n})}{%
\partial \theta _{n}}|_{\mathcal{\breve{\theta}}_{n}}\right) \left( \frac{%
\partial l_{n,i}(\theta _{n})}{\partial \theta _{n}^{\prime }}|_{\mathcal{%
\breve{\theta}}_{n}}\right) ,$ where $l_{n,i}(\theta _{n})$ is the
contribution to the reparametrized log-likelihood function, $l_{n}(\theta
_{n})$, by individual $i$. If $\widetilde{\mathcal{\theta }}_{n}\neq \theta
_{\ast }$ for all $\sigma ^{2}>0$, then $QLM(\widetilde{\mathcal{\theta }}%
_{n})$ is given by, cf. White (1994, p. 173):\vspace{-0.1in} 
\begin{eqnarray}
QLM(\widetilde{\mathcal{\theta }}_{n}) &=&N^{-1}\times \frac{\partial l_{n}(%
\widetilde{\mathcal{\theta }}_{n})}{\partial \theta _{n}^{\prime }}\overline{%
H}^{-1}(\widetilde{\mathcal{\theta }}_{n})A^{\prime }\times  \label{qlm} \\
&&(A\overline{H}^{-1}(\widetilde{\mathcal{\theta }}_{n})\overline{\mathcal{J}%
}(\widetilde{\mathcal{\theta }}_{n})\overline{H}^{-1}(\widetilde{\mathcal{%
\theta }}_{n})A^{\prime })^{-1}A\overline{H}^{-1}(\widetilde{\mathcal{\theta 
}}_{n})\frac{\partial l_{n}(\widetilde{\mathcal{\theta }}_{n})}{\partial
\theta _{n}}.  \notag
\end{eqnarray}%
If $H_{0}$ is true and $\mathcal{\theta }_{0,n}\neq \theta _{\ast }$ (for
all $\sigma ^{2}>0$), then $QLM(\widetilde{\mathcal{\theta }}_{n})\overset{d}%
{\rightarrow }\chi ^{2}(J).$ To test $H_{0}:$ $\rho =a$ when $\widetilde{%
\mathcal{\theta }}_{n}\neq \theta _{\ast }$ (for all $\sigma ^{2}>0$,) for
some known value of $a\in (-1,1]$, one can use $QLM(\widetilde{\mathcal{%
\theta }}_{n})$ in (\ref{qlm}) with $A=(1$ $\mathbf{0}^{\prime })$ and $%
\frac{\partial l_{n}(\widetilde{\mathcal{\theta }}_{n})}{\partial \theta _{n}%
}=A^{\prime }\frac{\partial l_{n}(\widetilde{\mathcal{\theta }}_{n})}{%
\partial \rho }$. Note also that, unlike the values of the\linebreak (Quasi)
Wald and Hausman test-statistics that use $\overline{H}^{-1}(\widehat{%
\mathcal{\theta }}_{n})$ (cf. White, 1994, p. 173), the value of $QLM(%
\widetilde{\mathcal{\theta }}_{n})$ remains the same when $\overline{H}^{-1}(%
\mathcal{\breve{\theta}}_{n})$ for some $\mathcal{\breve{\theta}}_{n}$ is
replaced by $adj(\overline{H}(\mathcal{\breve{\theta}}_{n}))$. Furthermore,
if $\widetilde{\mathcal{\theta }}_{n}\neq \theta _{\ast }$, then $\overline{%
\mathcal{J}}(\widetilde{\mathcal{\theta }}_{n})$ and p$\lim_{N\rightarrow
\infty }\overline{\mathcal{J}}(\widetilde{\mathcal{\theta }}_{n})$ are
positive definite.

If $\widetilde{\mathcal{\theta }}_{n}=\theta _{\ast }$ for some $\sigma
^{2}>0,$ $rk(\overline{H}(\widetilde{\mathcal{\theta }}_{n}))=\dim (%
\overline{H}(\widetilde{\mathcal{\theta }}_{n}))-1$ and hence $rk(adj(%
\overline{H}(\widetilde{\mathcal{\theta }}_{n})))=1.$ Furthermore, $%
\overline{H}(\widetilde{\mathcal{\theta }}_{n})_{i,j}=0$ iff $i=1$ and/or $%
j=1,$ $adj(\overline{H}(\widetilde{\mathcal{\theta }}_{n}))_{i,j}\neq 0$ iff 
$i=j=1,$ and $adj(\overline{H}(\widetilde{\mathcal{\theta }}_{n}))(1$ $%
\mathbf{0}^{\prime })^{\prime }\propto (1$ $\mathbf{0}^{\prime })^{\prime }.$
Hence if $\widetilde{\mathcal{\theta }}_{n}=\theta _{\ast }$ (for some $%
\sigma ^{2}>0$), the QLM test statistic in (\ref{qlm}) for $H_{0}:$ $\rho =a$
would be equal to $(\frac{\partial l_{n}(\mathcal{\theta }_{n})}{\partial r}%
|_{\theta _{\ast }})^{2}/(\tsum_{i=1}^{N}(\frac{\partial l_{n,i}(\mathcal{%
\theta }_{n})}{\partial r}|_{\theta _{\ast }})^{2})$. However, this test
statistic cannot be used, because $\frac{\partial l_{n,i}(\mathcal{\theta }%
_{n})}{\partial r}|_{\theta _{\ast }}=0$ a.s.\thinspace and p$%
\lim_{N\rightarrow \infty }N^{-1}\tsum_{i=1}^{N}(\frac{\partial l_{n,i}(%
\mathcal{\theta }_{n})}{\partial r}|_{\theta _{\ast }})^{2}=0$. More
generally, $N^{-1/2}A\,adj(\overline{H}(\theta _{\ast }))\frac{\partial
l_{n}(\mathcal{\theta }_{n})}{\partial \mathcal{\theta }_{n}}|_{\theta
_{\ast }}=0$ a.s. and p$\lim_{N\rightarrow \infty }[A\,adj(\overline{H}%
(\theta _{\ast }))\times $ $\overline{\mathcal{J}}(\theta _{\ast })adj(%
\overline{H}(\theta _{\ast }))A^{\prime }]=\mathbf{0.}$ To test $%
H_{0}:\nolinebreak $ $\nolinebreak A\mathcal{\theta }_{0,n}=a$ when $%
\widetilde{\mathcal{\theta }}_{n}=\theta _{\ast }$ (for some $\sigma ^{2}>0$%
), one can use the following QLM test statistic, cf. Bottai (2003):\vspace{%
-0.1in} 
\begin{eqnarray}
QLM(\widetilde{\mathcal{\theta }}_{n}) &=&N^{-1}\times \widetilde{S}^{\prime
}(\widetilde{\mathcal{\theta }}_{n})\widetilde{\mathcal{H}}^{-1}(\widetilde{%
\mathcal{\theta }}_{n})A^{\prime }\times  \label{qlm1} \\
&&(A\widetilde{\mathcal{H}}^{-1}(\widetilde{\mathcal{\theta }}_{n})%
\widetilde{\mathcal{J}}(\widetilde{\mathcal{\theta }}_{n})\widetilde{%
\mathcal{H}}^{-1}(\widetilde{\mathcal{\theta }}_{n})A^{\prime })^{-1}A%
\widetilde{\mathcal{H}}^{-1}(\widetilde{\mathcal{\theta }}_{n})\widetilde{S}(%
\widetilde{\mathcal{\theta }}_{n}),\vspace{-0.12in}  \notag
\end{eqnarray}%
with\pagebreak 
\begin{eqnarray*}
\widetilde{S}(\widetilde{\mathcal{\theta }}_{n})
&=&\sum\nolimits_{i=1}^{N}S_{i},\text{\quad }\widetilde{\mathcal{J}}(%
\widetilde{\mathcal{\theta }}_{n})=N^{-1}\sum%
\nolimits_{i=1}^{N}(S_{i}S_{i}^{\prime }),\medskip \\
S_{i} &=&(S_{i,1},S_{i,2}^{\prime })^{\prime },\text{\quad }S_{i,1}=\frac{1}{%
2}\frac{\partial ^{2}l_{n,i}}{\partial r_{n}^{2}}|_{\widetilde{\mathcal{%
\theta }}_{n}},\text{\quad }S_{i,2}=\frac{\partial l_{n,i}}{\partial d_{n}}%
|_{\widetilde{\mathcal{\theta }}_{n}},\vspace*{-0.35in}
\end{eqnarray*}%
$\vspace*{-0.25in}$and\vspace{-0.07in}%
\begin{eqnarray*}
\widetilde{\mathcal{H}}_{1,1} &=&\frac{2}{4!}E_{\widetilde{\mathcal{\theta }}%
_{n}}(\frac{\partial ^{4}l_{n}}{\partial r_{n}^{4}}|_{\widetilde{\mathcal{%
\theta }}_{n}}),\text{\quad }\widetilde{\mathcal{H}}_{1,2}^{\prime }=%
\widetilde{\mathcal{H}}_{2,1}=\medskip \frac{1}{2!}E_{\widetilde{\mathcal{%
\theta }}_{n}}(\frac{\partial ^{3}l_{n}}{\partial r_{n}^{2}\partial d_{n}}|_{%
\widetilde{\mathcal{\theta }}_{n}}), \\
\widetilde{\mathcal{H}}_{2,2} &=&\frac{2}{2!}E_{\widetilde{\mathcal{\theta }}%
_{n}}(\frac{\partial ^{2}l_{n}}{\partial d_{n}\partial d_{n}^{\prime }}|_{%
\widetilde{\mathcal{\theta }}_{n}}),\text{\quad }\widetilde{\mathcal{H}}(%
\widetilde{\mathcal{\theta }}_{n})=\left[ 
\begin{array}{cc}
\widetilde{\mathcal{H}}_{1,1} & \widetilde{\mathcal{H}}_{1,2} \\ 
\widetilde{\mathcal{H}}_{2,1} & \widetilde{\mathcal{H}}_{2,2}%
\end{array}%
\right] ,\vspace{-0.1in}
\end{eqnarray*}%
where we have used $l_{n}$ and $l_{n,i}$ as short for $l_{n}(\theta _{n})$
and $l_{n,i}(\theta _{n})$. In contrast, the Wald and Hausman
test-statistics that use $\overline{H}^{-1}(\widehat{\mathcal{\theta }}_{n})$
when $\widehat{\mathcal{\theta }}_{n}\neq \theta _{\ast }$ cannot be defined
when $\widehat{\mathcal{\theta }}_{n}=\theta _{\ast }.$ If $H_{0}$ is true
and $\mathcal{\theta }_{0,n}=\theta _{\ast }$ for some $\sigma ^{2}>0$, then 
$QLM(\theta _{\ast })\overset{d}{\rightarrow }\chi ^{2}(J).$ To test $H_{0}:$
$\rho =a$ when $\widetilde{\mathcal{\theta }}_{n}=\theta _{\ast }$, one can
use $QLM(\widetilde{\mathcal{\theta }}_{n})$ in (\ref{qlm1}) with $A=(1$ $%
\mathbf{0}^{\prime })$ and $\widetilde{S}(\widetilde{\mathcal{\theta }}%
_{n})=A^{\prime }(\sum\nolimits_{i=1}^{N}S_{i,1})$. It can easily be shown
that the test statistic $QLM(\widetilde{\mathcal{\theta }}_{n})$ for testing 
$H_{0}:$ $\rho =a$ given by (\ref{qlm}) and (\ref{qlm1}) is continuous at $%
\widetilde{\mathcal{\theta }}_{n}=\theta _{\ast }$ (for any $\sigma ^{2}>0$)
by using de l'H\^{o}pital's rule twice.\vspace{-0.06in}

\begin{theorem}
Under regularity conditions (A1)-(A7) given in the appendix, the Quasi LM
test for $H_{0}:$ $A\mathcal{\theta }_{0,n}=a,$ where $A_{1,.}=(1$ $\mathbf{0%
}^{\prime }),$ that is based on (\ref{qlm}) if $\widetilde{\mathcal{\theta }}%
_{n}\neq \theta _{\ast }$ for all $\sigma ^{2}>0,$ and on (\ref{qlm1}) if $%
\widetilde{\mathcal{\theta }}_{n}=\theta _{\ast }$ for some $\sigma ^{2}>0,$
has correct asymptotic size in a uniform sense.\vspace{-0.04in}
\end{theorem}

Confidence Sets (CSs) that have correct uniform asymptotic size can be
obtained by "inverting" the QLM test, i.e., $QLM(\widetilde{\mathcal{\theta }%
}_{n})$ given by (\ref{qlm}) and (\ref{qlm1}), or any of the other
aforementioned tests that have correct uniform asymptotic size. For
instance, a CS for $\rho $ of level $(1-\alpha )\ast 100\%$ constructed in
this way is the set of points $a\in {\mathbb{R}}$ for which the test fails
to reject $H_{0}:\rho =a$ at significance level $\alpha .$

When the data are i.i.d. and normal and $\rho $ is not local or equal to
one, the (Q)LM test that uses the expected Hessian has optimal local power
properties. Next we derive the power envelope of a centered version of $QLM(%
\widetilde{\mathcal{\theta }}_{n})$ for testing $H_{0}:$ $\rho =a=1-\kappa /%
\sqrt[4]{N}$, where $\kappa $ is a constant, viz. $QLM^{c}(\widetilde{%
\mathcal{\theta }}_{n}),$ when the data are i.i.d. and normal and $\theta
_{0}=\theta _{\ast }.$ Let $\overline{\mathcal{J}}^{c}\mathcal{(\theta }_{n}%
\mathcal{)}=N^{-1}\tsum_{i=1}^{N}\mathcal{J}_{i}^{c}\mathcal{(\theta }_{n}%
\mathcal{)}$ with $\mathcal{J}_{i}^{c}\mathcal{(\mathcal{\mathcal{\breve{%
\theta}}}}_{n}\mathcal{)}=(\frac{\partial l_{n,i}(\theta _{n})}{\partial
\theta _{n}}|_{\mathcal{\breve{\theta}}_{n}}-N^{-1}\tsum_{i=1}^{N}(\frac{%
\partial l_{n,i}(\theta _{n})}{\partial \theta _{n}}|_{\mathcal{\breve{\theta%
}}_{n}}))(\frac{\partial l_{n,i}(\theta _{n})}{\partial \theta _{n}^{\prime }%
}|_{\mathcal{\breve{\theta}}_{n}}-N^{-1}\tsum_{i=1}^{N}(\frac{\partial
l_{n,i}(\theta _{n})}{\partial \theta _{n}^{\prime }}|_{\mathcal{\breve{%
\theta}}_{n}}))$. If $\widetilde{\mathcal{\theta }}_{n}\neq \theta _{\ast },$
we have\vspace{-0.04in}%
\begin{eqnarray*}
QLM^{c}(\widetilde{\mathcal{\theta }}_{n}) &=&N^{-1}\times \frac{\partial
l_{n}(\widetilde{\mathcal{\theta }}_{n})}{\partial \rho }A\overline{H}^{-1}(%
\widetilde{\mathcal{\theta }}_{n})A^{\prime }\times \\
&&(A\overline{H}^{-1}(\widetilde{\mathcal{\theta }}_{n})\overline{\mathcal{J}%
}^{c}(\widetilde{\mathcal{\theta }}_{n})\overline{H}^{-1}(\widetilde{%
\mathcal{\theta }}_{n})A^{\prime })^{-1}A\overline{H}^{-1}(\widetilde{%
\mathcal{\theta }}_{n})A^{\prime }\frac{\partial l_{n}(\widetilde{\mathcal{%
\theta }}_{n})}{\partial \rho },\text{ }\vspace{-0.12in}\text{ }
\end{eqnarray*}%
where $A=(1$ $\mathbf{0}^{\prime }).$ If $\widetilde{\mathcal{\theta }}%
_{n}=\theta _{\ast },$ $QLM^{c}(\widetilde{\mathcal{\theta }}_{n})$ is
similar to (\ref{qlm1}) with $\widetilde{S}(\widetilde{\mathcal{\theta }}%
_{n})=A^{\prime }(\sum\nolimits_{i=1}^{N}S_{i,1})$ and $\widetilde{\mathcal{J%
}}(\widetilde{\mathcal{\theta }}_{n})$ replaced by $\widetilde{\mathcal{J}}%
^{c}(\widetilde{\mathcal{\theta }}_{n})=N^{-1}\sum\nolimits_{i=1}^{N}(S_{i}-%
\overline{S})(S_{i}-\overline{S})^{\prime },$ where $\overline{S}%
=N^{-1}\sum\nolimits_{i=1}^{N}S_{i}$.

If mean stationarity (\ref{mstat}) holds when $\left\vert \rho \right\vert
<1 $ and $Var(\Delta y_{i,t})/Var(y_{i,1})=o(N^{-1}),$ $i=1,2,...,N,$ then
only the full set of moment conditions based on differenced\linebreak data
matters asymptotically for identification of $\rho $ and there is no loss in
efficiency\linebreak when exploiting only the latter for estimation or
testing purposes; additional moment conditions involving levels of the data
are redundant (cf. Kruiniger, 2022, and Bun and Kleibergen, 2022). Note that 
$Var(\Delta y_{i,t})=O(1)$ for any $\rho \in (-1,1].$ We will investigate
the power of the $QLM^{c}(\widetilde{\mathcal{\theta }}_{n})$ tests under a
worst case scenario where $\theta _{0}=\theta _{\ast }$ and $%
1/Var(y_{i,1})=o(N^{-1})$ $i=1,2,...,N$ and consider a sequence of null
hypotheses that is local-to-unity: $H_{0}:$ $\rho =a=1-\kappa /\sqrt[4]{N}$
(cf. Bun and Kleibergen, 2022). In this scenario it is sufficient to focus
on the FE version of $QLM^{c}(\widetilde{\mathcal{\theta }}_{n}),$ viz. $%
QLM_{FE}^{c}(\widetilde{\mathcal{\theta }}_{n}).$

\begin{theorem}
When the data are i.i.d. and normal and $\theta _{0}=\theta _{\ast },$ (so
that TSH\ holds,) then the large sample distribution of $QLM_{FE}^{c}(%
\widetilde{\mathcal{\theta }}_{n})$ for testing $H_{0}:$ $\rho =a=1-\kappa /%
\sqrt[4]{N}$ is given by $\chi ^{2}($p$\lim_{N\rightarrow \infty
}c_{1}^{\prime }S^{-1}\overline{H}^{-1}A^{\prime }(A\overline{H}^{-1}%
\overline{\mathcal{J}}^{c}\overline{H}^{-1}A^{\prime })^{-1}A\overline{H}%
^{-1}S^{-1}c_{1}\kappa ^{4},1)$ where $c_{1}$ and $S$ are defined in the
proof in the appendix.
\end{theorem}

The quartic root rate in the sequence of hypotheses is related to the fact
that $\rho $ is only second-order identified when $\theta _{0}=\theta _{\ast
}$. When we impose TSH on $l_{n}(\theta _{n}),$ we obtain the following
result:

\begin{theorem}
When the data are i.i.d. and normal, $\theta _{0}=\theta _{\ast }$ and TSH
has been imposed on $l_{n}(\theta _{n})$, then the large sample distribution
of $QLM_{FE}^{c}(\widetilde{\mathcal{\theta }}_{n})$ for testing $H_{0}:$ $%
\rho =a=1-\kappa /\sqrt[4]{N}$ is given by $\chi ^{2}(\frac{(2T-3)T(T-1)(T-2)%
}{72}\kappa ^{4},1).$
\end{theorem}

We will obtain the maximal attainable power (MAP) curve for testing $H_{0}:$ 
$\rho =a=1-\kappa /\sqrt[4]{N}$ by first considering the asymptotic
distribution of an GMM-Anderson-Rubin statistic which tests $H_{0}$ by using
all the moment conditions based on first-differences of the data whilst (the
true value of) $\rho =1.$ This GMM-AR statistic is given by%
\begin{equation*}
GMM-AR(\rho )=Nm(\rho )^{\prime }[\widehat{V}_{mm}(\rho )]^{-1}m(\rho )%
\vspace{-0.12in}
\end{equation*}%
with$\vspace{-0.12in}$ 
\begin{equation}
m(r)=N^{-1}\sum\nolimits_{i=1}^{N}\left[ P\times vech\left( D_{r}(\Delta
y_{i}(\Delta y_{i})^{\prime })D_{r}^{\prime }\right) \right]  \label{momc}
\end{equation}%
where $D_{r}$ is a $(T-1)\times (T-1)$ band matrix with $(D_{r})_{i,i}=1$
and $(D_{r})_{i+1,i}=-r$ for $i=1,2,...,T-2,($and $T-1)$ and $%
(D_{r})_{i,j}=0 $ elsewhere; $P=(0$ $g$ $I_{p})$ is a $p\times \frac{1}{2}%
(T-1)T$ matrix where $p=\frac{1}{2}T(T-1)-2$ and $g$ is a $p$-vector such
that $(1,-1,$ $g^{\prime })^{\prime }=vech(D_{r}D_{r}^{\prime })$; and $%
\widehat{V}_{mm}(\rho )$ is the Eicker-White estimator of the covariance
matrix of $m(\rho ).$

\begin{theorem}
When the data are i.i.d. and normal, $\theta _{0}=\theta _{\ast }$ and TSH\
is exploited by the testing procedure, then the large sample distribution of 
$GMM-AR(\rho )$ for testing\linebreak $H_{0}:$ $\rho =a=1-\kappa /\sqrt[4]{N}
$ is given by $\chi ^{2}(\frac{(2T-3)T(T-1)(T-2)}{72}\kappa ^{4},\frac{1}{2}%
T(T-1)-2).$
\end{theorem}

We obtain the MAP curve for testing $H_{0}:$ $\rho =a=1-\kappa /\sqrt[4]{N}$
by deriving the asymptotic distribution of an GMM-Anderson-Rubin statistic
which tests $H_{0}$ by using the (infeasible) weighted average of the moment
conditions in $E(m(\rho ))=0_{p}$ (with $m(r)$ given in (\ref{momc})) that
leads to the largest value of the non-centrality parameter of this
distribution whilst (the true value of) $\rho =1$.

\begin{theorem}
When the data are i.i.d. and normal, $\theta _{0}=\theta _{\ast }$ and TSH\
is exploited by the testing procedure, then the maximal attainable power
curve for testing $H_{0}:$ $\rho =a=1-\kappa /\sqrt[4]{N}$ is given by $\chi
^{2}(\frac{(2T-3)T(T-1)(T-2)}{72}\kappa ^{4},1).$
\end{theorem}

\begin{corollary}
When the data are i.i.d. and normal, $\theta _{0}=\theta _{\ast }$ and TSH\
has been imposed on $l_{n}(\theta _{n})$, then the large sample distribution
of $QLM_{FE}^{c}(\widetilde{\mathcal{\theta }}_{n})$ for testing $H_{0}:$ $%
\rho =a=1-\kappa /\sqrt[4]{N}$ attains the maximal attainable power curve
for testing $H_{0}.$
\end{corollary}

This result implies that, like the KLM-statistic, the centered LM-statistic $%
QLM^{c}(\widetilde{\mathcal{\theta }}_{n})$ is efficient both when $\rho $
is less than one and when $\rho $ is equal to one.

To test $H_{0}:$ $\rho =1$ one could also use a Wald test based on $\sqrt{N}(%
\widehat{\rho }-1)^{2}$ where $\widehat{\rho }$ is the REQMLE or FEQMLE of $%
\rho .$ Under $H_{0}\ $we have $\sqrt{N}(\widehat{\rho }-1)^{2}\overset{d}{%
\rightarrow }\breve{Z}_{1}\mathbf{1}\{\breve{Z}_{1}>0\},$ cf. Theorem 1.
Recall that $\breve{Z}_{N}=(\breve{Z}_{1,N},\breve{Z}_{2,N}^{\prime
})^{\prime }=-\mathcal{H}^{-1}U_{2,N}\overset{d}{\rightarrow }\breve{Z}$
with $U_{2,N}=(\frac{1}{2!}N^{-1/2}\frac{\partial ^{2}l_{n}}{\partial
r_{n}^{2}}|_{\theta _{\ast }},N^{-1/2}\frac{\partial l_{n}}{\partial
d_{n}^{\prime }}|_{\theta _{\ast }})^{\prime }$ and $\breve{Z}=(\breve{Z}%
_{1},\breve{Z}_{2}^{\prime })^{\prime }\sim N(0,\mathcal{H}^{-1}\mathcal{IH}%
^{-1}).$ When the data are i.i.d. non-normal or i.h.d., one can bootstrap
the distribution of $U_{2,N}$. To do this, one can make use of the fact that
under $H_{0}$, $\varepsilon _{i}=y_{i}-y_{i,-1}$ for $i=1,...,N.$\vspace{%
-0.15in}

\section{The finite sample performance of the QLM\ tests\protect\vspace{%
-0.1in}}

In this section we investigate through Monte Carlo simulations the empirical
size and power properties of QLM-tests for testing a simple hypothesis of
the type $H_{0}:$ $\rho =a$, namely $QLM(\mathcal{\rho }),$ that are based
on the RE and FE\ likelihood functions for various panel AR(1) models
without covariates. The data were generated using the panel AR(1) model
given in (\ref{mdl}). We study how the properties of these tests are
affected if we change (1) the value of $\rho ,$ (2) the distributions of the
initial conditions $v_{i,1}=y_{i,1}-\mu _{i}$, (3) the distributions of the
idiosyncratic errors (the $\varepsilon _{i,t}$) and/or (4) the ratio of the
variances of the error components. We conducted the simulation experiments
for $(T,N)=(4,100),$ $(9,100),$ $(4,250)$ or $(9,250).$ The nominal size of
the tests was $0.05$ and for each scenario the number of replications was
2500.

We calculated the empirical size of the tests for $\rho =0.2,$ $0.5,$ $0.8,$ 
$0.9,$ $0.95,$ $0.98$ or $0.99$ and we calculated the empirical power of the
tests for $H_{0}:\rho =0.8$ when $\rho =0.5,$ $0.6,$ $0.7,$ $0.9,$ $0.95$ or 
$0.99.$

In all simulation experiments the individual effects, the $\mu _{i}$, were
i.i.d. $N(0,\sigma _{\mu }^{2})$ with $\sigma _{\mu }^{2}=1$ or $25.$ The
errors, the $\varepsilon _{i,t}$, were either i.i.d. $N(0,1)$ or i.i.d. $%
(\chi ^{2}(1)-1)/\sqrt{2}.$ Note that in both cases $Var(\varepsilon
_{i,t})=1$ for $i=1,...,N$ and $t=2,...,T$.

In order to assess how the assumptions with respect to $y_{i,1}-\mu _{i}$, $%
i=1,...,N,$ affect the finite sample properties of the tests, we conducted
three different kinds of experiments: in one set, labeled NS-Normal, the
initial observations are non-stationary, i.e., $y_{i,1}-\mu _{i}=0$, $%
i=1,...,N,$ and the $\varepsilon _{i,t}\sim N(0,1)$, whereas in the other
two sets, labeled S-Normal and S-ChiSq., respectively, the initial
observations are drawn from stationary distributions, i.e., either $%
(y_{i,1}-\mu _{i})\sim N(0,1/(1-\rho ^{2}))$ when the $\varepsilon
_{i,t}\sim N(0,1)$ or $(y_{i,1}-\mu _{i})\sim (\chi ^{2}(1)-1)/\sqrt{%
2(1-\rho ^{2})}$ when the $\varepsilon _{i,t}\sim (\chi ^{2}(1)-1)/\sqrt{2}$%
. Note that in design NS-Normal the data are still mean stationary, i.e., $%
E(y_{i,1}-\mu _{i})=0$ and $E(\mu _{i}(y_{i,1}-\mu _{i}))=0$, and that in
all designs $E(y_{i,t}-y_{i,t-1})=0$.

In the cases of the RE and FE models, $(1-\rho )\mu _{i}+\varepsilon _{i}$
is decomposed as $(1-\rho )\pi y_{i,1}+(1-\rho )v_{i}+\varepsilon
_{i}=(1-\rho )\pi y_{i,1}+u_{i}$ with $\pi =1$ for the FE case. In the
experiments we imposed homoskedasticity on the likelihood functions and
added the restrictions $\sigma ^{2}>0$ and $\widetilde{\sigma }_{v}^{2}\geq
0 $ or the restrictions $\sigma ^{2}>0$ and $(T-1)\widetilde{\sigma }%
_{v}^{2}+\sigma ^{2}>0$ in case the restriction $\widetilde{\sigma }%
_{v}^{2}\geq 0$ was binding to ensure that the estimates of $%
E(u_{i}u_{i}^{\prime })$ were positive definite.

We allowed for time effects by subtracting cross-sectional averages from the
data.

Note that the QML estimators suffer from a weak moment conditions problem
when $\rho $ is close to one.

Tables 1-12 report the simulation results in terms of the relative rejection
frequenties of the tests. Tables 1-4, 9, 10 report results concerning the
empirical size of the tests whereas tables 5-8, 11, 12 report results
concerning the empirical power of the tests. The tables also differ with
respect to the the value of $\sigma _{\mu }^{2}$: tables 1-8 correspond to $%
\sigma _{\mu }^{2}=1$ and report results for both RE and FE versions of the
QLM tests whereas tables 9-12 correspond to $\sigma _{\mu }^{2}=25$ and
report results for the RE version of the QLM\ test only because changes in
the value of $\sigma _{\mu }^{2}$ do not affect the FE affects version of
the QLM test by construction. If the tests have correct size (i.e., 0.05),
then the standard errors of the estimates of the empirical size are $\sqrt{%
0.05\ast (1-0.05)}/\sqrt{2500}\approx 0.0044.$

Inspection of the results in tables 1-4, 9 and 10 leads to the following
conclusions regarding the empirical size of the QLM tests:

\begin{enumerate}
\item Using a 5\% significance level, we would not be able to reject the
hypothesis that the tests have correct size for the various scenarios with $%
\sigma _{\mu }^{2}=1.$ Only 8 out of 168 estimates of the empirical size lie
outside the acceptance region $(0.0412,0.0588).$ The most extreme estimates
of the empirical sizes are $0.0628$ and $\ 0.0380.$ If we restrict attention
to $N=250$ and $\rho =0.95,$ $0.98$ or $0.99,$ then 2 out of 36 estimates of
the empirical size lie outside the acceptance region $(0.0412,0.0588).$

\item Using a 5\% significance level, we would not be able to reject the
hypothesis that the tests have correct size for the various scenarios with $%
\sigma _{\mu }^{2}=25.$ Only 5 out of 84 estimates of the empirical size lie
outside the acceptance region $(0.0412,0.0588).$ The most extreme estimates
of the empirical sizes are $0.0664$ and $\ 0.0644.$ If we restrict attention
to $N=250$ and $\rho =0.95,$ $0.98$ or $0.99,$ then 1 out of 18 estimates of
the empirical size lies outside the acceptance region $(0.0412,0.0588).$
\end{enumerate}

Inspection of the results in tables 5-8, 11 and 12 leads to the following
conclusions regarding the empirical power of the QLM tests:

\begin{enumerate}
\item[3.] The power of the tests increases in $N$ and $T.$

\item[4.] The power of the RE version of the QLM\ test is higher than the
power of the FE version of the QLM test unless the initial conditions $%
v_{i,1}=y_{i,1}-\mu _{i}$ are zero (NS) in which case the power is the same
for both versions of the QLM tests.

\item[5.] The power curves for testing $H_{0}:\rho =0.8$ are asymmetric
around $\rho =0.8$, i.e., the QLM\ tests have more power against an
alternative below $\rho =0.8$ than against an equidistant alternative above $%
\rho =0.8.$\vspace{-0.16in}
\end{enumerate}

\section{Concluding remarks}

In this paper we proposed new ML based inference methods for panel AR models
with arbitrary initial conditions and heteroskedasticity and possibly
additional regressors that are robust to the strength of identification.
Specifically, we showed that (Quasi) LM tests and CSs that use the expected
Hessian rather than the observed Hessian of the RE or the FE\ log-likelihood
function have correct asymptotic size in a uniform sense. We also derived
the power envelope of a FE version of such an LM test for testing $H_{0}:$ $%
\rho =a=1-\kappa /\sqrt[4]{N}$ when the average information matrix is
estimated by a centered OPG\ estimator and the model is only second-order
identified, and showed that it coincides with the maximal attainable power
curve for testing $H_{0}:$ $\rho =a=1-\kappa /\sqrt[4]{N}$ in the worst case
setting. In a Monte Carlo study that included a variety of experiments we
found that these (Quasi) LM tests have correct empirical size and good
empirical power properties.

None of the existing ML based inference methods for dynamic panel data
models have correct uniform asymptotic size close to the point in the
parameter space at which such a model is only second-order identified and
therefore the methods proposed in this paper will result in more reliable
inference. The proposed methods can be adapted to generalizations of the
panel AR(1) model that was considered in this paper including dynamic panel
models with a factor structure and panel VAR models. Similarly, Quasi LM\
test that have correct uniform asymptotic size can also be developed for
other multi-parameter models that are only second-order identified at some
point in the parameter space, such as the examples mentioned in e.g.
Rotnitzky et. al. (2000) and Dovonon and Hall (2018).

An issue that requires additional study is the effect of the choice of the
estimator for the average information matrix on the power of the QLM tests.
\newpage

\appendix 

\section{Proofs$\protect\vspace{-0.1in}$}

\textbf{\noindent Regularity conditions:}$\smallskip $

Let $f(y_{i}^{+};\theta )$ denote the density of the random vector $%
y_{i}^{+} $ so that $l_{i}(\theta )=\ln f(y_{i}^{+};\theta ).$ Let $p=\dim
(\theta ).$ We can write $\theta $ as $(\theta _{1},...,\theta _{p})^{\prime
}.$ For any $1\times p$ vector $a=(a_{1},...,a_{p}),$ let $%
l_{i}^{(a)}(\theta )$ denote $\partial ^{a.}l_{i}(\theta )/\partial
^{a_{1}}\theta _{1}\partial ^{a_{2}}\theta _{2}...\partial ^{a_{p}}\theta
_{p}$ where $a.=\tsum\nolimits_{k=1}^{p}a_{k}.$ Let $l^{(a)}(\theta
)=\tsum\nolimits_{i=1}^{N}l_{i}^{(a)}(\theta )$ and define $%
f^{(a)}(y_{i}^{+};\theta )$ similarly. Let w.p.1 denote "with probability
1". Let $S_{k}(\theta )=\partial l(\theta )/\partial \theta _{k},$ $1\leq
k\leq p,$ and $S_{k}=S_{k}(\theta _{\ast }).$ Let $S_{1}^{(q+j)},$ $j=0,1,$
denote $\partial ^{q+j}l(\theta )/\partial ^{q+j}\theta _{1}|_{\theta _{\ast
}}.$ Let $q\in 
\mathbb{N}
$ be such that conditions (B1$^{\prime }$)-(B3$^{\prime }$) given in K2013
hold, i.e., $\theta _{1}$ is $q$-th order identified. Then the regularity
conditions for the case of i.h.d. data are given by (cf. Bottai, 2003):

\begin{description}
\item[(A1$^{\prime }$)] $\theta _{\ast }\in \Theta $, a compact subset of $%
\mathbb{R}
^{p}$ that contains an open neighbourhood $\mathcal{N}$ of $\theta _{\ast }$.%
\vspace{-0.06in}

\item[(A2$^{\prime }$)] Distinct values of $\theta $ in $\Theta $ correspond
to distinct probability distributions.\vspace{-0.06in}

\item[(A3$^{\prime }$)] W.p.1 (under $\theta _{\ast }$), $f(y_{i}^{+};\theta
)>0$ for all $\theta \in \mathcal{N}$.\vspace{-0.06in}

\item[(A4$^{\prime }$)] There exist $B_{i}(y_{i}^{+})$ such that $\left\vert
l_{i}(\theta )\right\vert \leq B_{i}(y_{i}^{+})$ for all $\theta \in \Theta $
and $E\left\vert B_{i}(y_{i}^{+})\right\vert ^{1+\xi }<\infty $ for some $%
\xi >0$ and all $i\in \{1,...,N\}.$\vspace{-0.06in}

\item[(A5$^{\prime }$)] For all $\theta $ in $\mathcal{N}$ and all $a$ with $%
1\leq a.\leq 2q+1$, the derivatives $f^{(a)}(y_{i}^{+};\theta )$ and $%
l_{i}^{(a)}(\theta )$ exist w.p.1 and there exist $B_{i}(y_{i}^{+})$ such
that $\left\vert l_{i}^{(a)}(\theta )\right\vert \leq B_{i}(y_{i}^{+})$ for
all $\theta \in \mathcal{N}$ and $E\left\vert B_{i}(y_{i}^{+})\right\vert
^{1+\xi }<\infty $ for some $\xi $ and all $i\in \{1,...,N\}.$ Furthermore
(for all $a$ with $1\leq a.\leq 2q+1$), $\int \sup_{\theta \in \mathcal{N}%
}\left\vert f^{(a)}(y_{i}^{+};\theta )\right\vert dy_{i}<\infty $ , $\int
\sup_{\theta \in \mathcal{N}}[\{f^{(a)}(y_{i}^{+};\theta
)\}^{2}/f(y_{i}^{+};\theta )]dy_{i}<\infty ,$ and $\int \sup_{\theta \in 
\mathcal{N}}\{\left\vert l_{i}^{(a)}(\theta )\right\vert ^{j}f^{(a^{\prime
})}(y_{i}^{+};\theta )\}dy_{i}<\infty $ for $j=1,2$ and all $a^{\prime }$
with $0\leq a^{\prime }.\leq 2q+1,$ where $dy_{i}$ is short for $%
dy_{i,1}dy_{i,2},...,dy_{i,T}$\vspace{-0.06in}

\item[(A6a$^{\prime }$)] For all $a$ with $1\leq a.\leq 2q+1,$ there exist $%
\xi (a)>2$ such that $\sup_{\theta \in \mathcal{N}}E_{\theta }\left\vert
l_{i}^{(a)}(\theta ^{\prime })\right\vert ^{\xi (a)}$ $<\infty $ for all $%
\theta ^{\prime }\in \mathcal{N}$ and all $i\in \{1,...,N\}.$\vspace{-0.06in}

\item[(A6b$^{\prime }$)] When $a.=2q+1$ there exists $\varpi >0$ and some
function $h(.)$ satisfying $E\left\vert h(y_{i}^{+})\right\vert ^{\xi }$ $%
<\infty $ for some $\xi >2$ and all $i\in \{1,...,N\}$, such that for any $%
\theta $ and $\theta ^{\prime }$ in $\mathcal{N}$, w.p.1, $\left\vert
l^{(a)}(\theta )-l^{(a)}(\theta ^{\prime })\right\vert \leq \left\Vert
\theta -\theta ^{\prime }\right\Vert ^{\varpi }h(y_{i}^{+}),$ where $%
\left\Vert \theta \right\Vert =\left( \tsum\nolimits_{k=1}^{p}\theta
_{k}^{2}\right) ^{1/2}.$\vspace{-0.06in}

\item[(A7$^{\prime }$)] Conditions (B1$^{\prime }$)-(B3$^{\prime }$) given
in K2013 hold and w.p.1, $S_{2},...,S_{p}$ are linearly independent.
Furthermore, for each $\theta \neq \theta _{\ast }$, $S_{1}(\theta )\neq 0$
with positive probability.\vspace{-0.06in}
\end{description}

Conditions (A1)-(A7) alluded to in our theorem 4 are equal to conditions (A1$%
^{\prime }$)-(A7$^{\prime }$), respectively, with $p=T+2$, $q=2$ and $\theta
_{\ast }=(\rho _{\ast },\delta _{\ast }^{\prime })^{\prime }=(1,\delta
_{\ast }^{\prime })^{\prime }$ and with $\theta $ and $l_{i}(\theta )$
replaced by $\theta _{n}=(r_{n},d_{n}^{\prime })^{\prime }$ and $%
l_{n,i}(\theta _{n}),$ respectively. In this case (B1$^{\prime }$)-(B3$%
^{\prime }$) given in K2013 are equal to conditions (B1)-(B3) given above
theorem 1. Conditions (A1)-(A7) are satisfied when the data are i.i.d. and
normal but in general have to be checked.$\smallskip \smallskip $

\textbf{\noindent Proof of theorem 4:}$\smallskip $

We first prove that if (A1)-(A7) hold, the restricted QMLE $\widetilde{%
\mathcal{\theta }}_{n}=\widetilde{\mathcal{\theta }}_{n,N}$ that satisfies $A%
\widetilde{\mathcal{\theta }}_{n,N}=a_{N},$ where $A_{1,.}=(1$ $\mathbf{0}%
^{\prime }),$ is root-$N$ consistent under the parameter sequence $\mathcal{%
\theta }_{0,n,N}\linebreak $with $\mathcal{\theta }_{0,n,N}\rightarrow 
\mathcal{\theta }_{\ast }$ and $A\mathcal{\theta }_{0,n,N}=a_{N}$ (so that $A%
\mathcal{\theta }_{\ast }=a=\lim_{N\rightarrow \infty }a_{N}$). Consistency
of $\widetilde{\mathcal{\theta }}_{n,N}$ follows from Theorem 2.1 in NMcF.
The proof is similar to that of consistency of $\widehat{\theta }_{n}$ given
in K2013wp. Let $\mathcal{\ddot{\theta}}_{n,N}=(a_{N},\mathcal{\ddot{\delta}}%
_{n,N}^{\prime })^{\prime }$ be the QMLE that maximizes $l_{n}(\theta _{n,N})
$ subject to $r_{n,N}=a_{N}$. One can show that $\mathcal{\ddot{\delta}}%
_{n,N}-\mathcal{\delta }_{n,N}=O_{p}(N^{-1/2})$ by expanding the likeli-$%
\linebreak $hood equations for $\delta $ corresponding to $l_{n}(\theta
_{n,N})$ with $r_{n,N}=a_{N}$. Following Davidson and MacKinnon (1993, pp.
276-277), one can similarly show that $\widetilde{\mathcal{\theta }}_{n,N}-%
\mathcal{\theta }_{0,n,N}=O_{p}(N^{-1/2})$.

Let $\overline{S}_{N,i}(\widetilde{\mathcal{\theta }}_{n,N};\mathcal{\theta }%
_{0,n,N})=(A\overline{H}^{-1}(\widetilde{\mathcal{\theta }}_{n,N})\overline{%
\mathcal{J}}(\widetilde{\mathcal{\theta }}_{n,N})\overline{H}^{-1}(%
\widetilde{\mathcal{\theta }}_{n,N})A^{\prime })^{-1/2}A\overline{H}^{-1}(%
\widetilde{\mathcal{\theta }}_{n,N})\frac{\partial \widetilde{l}_{n,i}(%
\widetilde{\mathcal{\theta }}_{n,N})}{\partial \theta _{n}}\linebreak $when $%
\widetilde{\mathcal{\theta }}_{n,N}\neq \theta _{\ast },$ $\mathcal{\theta }%
_{0,n,N}\neq \theta _{\ast }$ (for all$\,\sigma ^{2}>0$), $A\mathcal{\theta }%
_{0,n,N}=a_{N},$ and $\mathcal{\theta }_{0,n,N}\in \Theta .$ For any $\{%
\widetilde{\mathcal{\theta }}_{n,N}\}$ such that $\widetilde{\mathcal{\theta 
}}_{n,N}\rightarrow \mathcal{\theta }_{\ast },$ the limit of $\overline{QLM}(%
\widetilde{\mathcal{\theta }}_{n,N};\mathcal{\theta }_{0,n,N})\equiv
(N^{-1/2}\tsum_{i=1}^{N}\overline{S}_{N,i}(\widetilde{\mathcal{\theta }}%
_{n,N};\mathcal{\theta }_{0,n,N}))^{\prime }\times $ $(N^{-1/2}%
\tsum_{i=1}^{N}\overline{S}_{N,i}(\widetilde{\mathcal{\theta }}_{n,N};%
\mathcal{\theta }_{0,n,N}))$ when $\widetilde{\mathcal{\theta }}_{n,N}(=%
\widetilde{\mathcal{\theta }}_{n})\rightarrow \mathcal{\theta }_{\ast }$
(for any $\sigma ^{2}>0$) and $\mathcal{\theta }_{0,n,N}\rightarrow \mathcal{%
\theta }_{\ast }$ (for any $\sigma ^{2}>0$), while $A\mathcal{\theta }%
_{0,n,N}=a_{N}$, can be obtained by applying de l'H\^{o}pital's rule.

Following Davidson and MacKinnon (1993, pp. 276-277), we obtain $N^{1/2}(%
\widetilde{\mathcal{\theta }}_{n}-\mathcal{\theta }_{0,n,N})\overset{asy}{=}%
\linebreak -\overline{H}^{-1}(I-A^{\prime }(A\overline{H}^{-1}A^{\prime
})^{-1}A\overline{H}^{-1})N^{-1/2}\frac{\partial l_{n}(\mathcal{\theta }%
_{0,n,N})}{\partial \theta _{n}}$ where $\overline{H}\mathcal{=}\overline{H}(%
\mathcal{\theta }_{0,n,N})$ and $\mathcal{\theta }_{0,n,N}\neq \theta _{\ast
}.$ We also have $N^{-1/2}\frac{\partial l_{n}(\widetilde{\mathcal{\theta }}%
_{n})}{\partial \theta _{n}}\overset{asy}{=}N^{-1/2}\frac{\partial l_{n}(%
\mathcal{\theta }_{0,n,N})}{\partial \theta _{n}}+\overline{H}N^{1/2}(%
\widetilde{\mathcal{\theta }}_{n}-\mathcal{\theta }_{0,n,N}).$ Hence $%
N^{-1/2}\frac{\partial l_{n}(\widetilde{\mathcal{\theta }}_{n})}{\partial
\theta _{n}}\overset{asy}{=}\linebreak A^{\prime }(A\overline{H}%
^{-1}A^{\prime })^{-1}A\overline{H}^{-1}N^{-1/2}\frac{\partial l_{n}(%
\mathcal{\theta }_{0,n,N})}{\partial \theta _{n}}$ and $(A\overline{H}^{-1}(%
\widetilde{\mathcal{\theta }}_{n})\overline{\mathcal{J}}(\widetilde{\mathcal{%
\theta }}_{n})\overline{H}^{-1}(\widetilde{\mathcal{\theta }}_{n})A^{\prime
})^{-1/2}A\overline{H}^{-1}(\widetilde{\mathcal{\theta }}_{n})\times $%
\linebreak $N^{-1/2}\frac{\partial l_{n}(\widetilde{\mathcal{\theta }}_{n})}{%
\partial \theta _{n}}\overset{asy}{=}(A\overline{H}^{-1}(\mathcal{\theta }%
_{0,n,N})E_{\mathcal{\theta }_{0,n,N}}\left( \overline{\mathcal{J}}(\mathcal{%
\theta }_{0,n,N})\right) \overline{H}^{-1}(\mathcal{\theta }%
_{0,n,N})A^{\prime })^{-1/2}A\overline{H}^{-1}(\mathcal{\theta }%
_{0,n,N})\times \linebreak N^{-1/2}\frac{\partial l_{n}(\mathcal{\theta }%
_{0,n,N})}{\partial \theta _{n}}$ under the parameter sequence $\mathcal{%
\theta }_{0,n,N}$ with $\mathcal{\theta }_{0,n,N}\rightarrow \mathcal{\theta 
}_{\ast }$, $A\mathcal{\theta }_{0,n,N}=a_{N}$ and $\mathcal{\theta }%
_{0,n,N}\neq \theta _{\ast },$ where we have used that p$\lim_{N\rightarrow
\infty }[\overline{\mathcal{J}}(\widetilde{\mathcal{\theta }}_{n})-E_{%
\mathcal{\theta }_{0,n,N}}\left( \overline{\mathcal{J}}(\mathcal{\theta }%
_{0,n,N})\right) ]=0$ under $\mathcal{\theta }_{0,n,N}\rightarrow \mathcal{%
\theta }_{\ast }$ and the fact that $\overline{\mathcal{J}}(\widetilde{%
\mathcal{\theta }}_{n})$ and $E_{\mathcal{\theta }_{0,n,N}}\left( \overline{%
\mathcal{J}}(\mathcal{\theta }_{0,n,N})\right) $ are positive definite when $%
\widetilde{\mathcal{\theta }}_{n,N}\neq \theta _{\ast }$ and $\mathcal{%
\theta }_{0,n,N}\neq \theta _{\ast }$.

Let $S_{N,i}(\mathcal{\theta }_{n})=A\overline{H}^{-1}(\mathcal{\theta }_{n})%
\frac{\partial l_{n,i}(\mathcal{\theta }_{n})}{\partial \theta _{n}}$, $%
S_{N,i}=S_{N,i}(\mathcal{\theta }_{0,n,N})$ and $\mathcal{\theta }%
_{0,n,N}\rightarrow \mathcal{\theta }_{\ast }\in \Theta $ with $A\mathcal{%
\theta }_{0,n,N}=a_{N}$ and $\mathcal{\theta }_{0,n,N}\neq \theta _{\ast }.$
Under conditions (A5)-(A6) and similarly to Bottai (2003), $E_{\mathcal{%
\theta }_{0,n,N}}(S_{N,i})=0,$ $Var_{\mathcal{\theta }_{0,n,N}}(S_{N,i})=A%
\overline{H}^{-1}(\mathcal{\theta }_{0,n,N})E_{\mathcal{\theta }_{0,n,N}}(%
\mathcal{J}_{i}(\mathcal{\theta }_{0,n,N}))\overline{H}^{-1}(\mathcal{\theta 
}_{0,n,N})A^{\prime }$ and$\linebreak \sup_{i}\sup_{\mathcal{\theta }_{n}\in 
\mathcal{N}}E_{\mathcal{\theta }_{0,n,N}}(\left\vert \lambda ^{\prime
}S_{N,i}\right\vert ^{\varsigma })<\infty $ for some $\varsigma >2$ and for
all $\lambda \in 
\mathbb{R}
^{J}$ where $\mathcal{N}\subset \Theta $ is an open neighbourhood around $%
\mathcal{\theta }_{\ast }$. We also have $N^{-1}\tsum_{i=1}^{N}Var_{\mathcal{%
\theta }_{0,n,N}}(\lambda ^{\prime }S_{N,i})>0$ uniformly in $N$ for all $%
\lambda \in 
\mathbb{R}
^{J}\backslash \{\mathbf{0}\}.$ Thus the Lyapunov conditions are satisfied
and by (a multivariate version of) Lindeberg's CLT for triangular arrays, $%
(\tsum_{i=1}^{N}Var_{\mathcal{\theta }_{0,n,N}}(S_{N,i}))^{-1/2}\times
\linebreak \tsum_{i=1}^{N}S_{N,i}$ converges under the parameter sequence $%
\mathcal{\theta }_{0,n,N}$ to $N(0,I_{J}).$ It follows that $(A\overline{H}%
^{-1}(\widetilde{\mathcal{\theta }}_{n})\overline{\mathcal{J}}(\widetilde{%
\mathcal{\theta }}_{n})\overline{H}^{-1}(\widetilde{\mathcal{\theta }}%
_{n})A^{\prime })^{-1/2}A\overline{H}^{-1}(\widetilde{\mathcal{\theta }}%
_{n})N^{-1/2}\frac{\partial l_{n}(\widetilde{\mathcal{\theta }}_{n})}{%
\partial \theta _{n}}\overset{d}{\rightarrow }N(0,I_{J})$ and $QLM(%
\widetilde{\theta }_{n})\overset{d}{\rightarrow }\chi ^{2}(J)$ under the
parameter sequence $\mathcal{\theta }_{0,n,N}$ with $\mathcal{\theta }%
_{0,n,N}\rightarrow \mathcal{\theta }_{\ast }$, $A\mathcal{\theta }%
_{0,n,N}=a_{N}$ and $\mathcal{\theta }_{0,n,N}\neq \theta _{\ast }.$

Next let $\widetilde{S}_{N,i}(\mathcal{\theta }_{n})=A\widetilde{\mathcal{H}}%
^{-1}(\mathcal{\theta }_{n})\widetilde{S}(\mathcal{\theta }_{n})$ and $%
\widetilde{S}_{N,i}=\widetilde{S}_{N,i}(\mathcal{\theta }_{\ast }).$ Under
condition (A7) and similarly to Bottai (2003), we can also show that $%
(\tsum_{i=1}^{N}Var_{\mathcal{\theta }_{0,n,N}}(\widetilde{S}%
_{N,i}))^{-1/2}\tsum_{i=1}^{N}\widetilde{S}_{N,i}\overset{d}{\rightarrow }%
N(0,I_{J})$ and $QLM(\mathcal{\theta }_{\ast })\overset{d}{\rightarrow }\chi
^{2}(J)$ when $\mathcal{\theta }_{0,n,N}=\mathcal{\theta }_{\ast }.$ We
conclude that$\linebreak \lim_{N\rightarrow \infty }\sup_{\theta _{0}\in 
\mathcal{N}}\left\vert \Pr_{\theta _{0}}\{QLM(\widetilde{\theta }_{n})>\chi
_{J,\alpha }^{2}\}-\alpha \right\vert =0.$\quad $\square \medskip $

\textbf{\noindent Proof of theorem 5:}$\smallskip $

We first derive $\frac{\partial l}{\partial \theta _{0}},$ where $%
l=l_{FE}(\theta _{0})$ with $\theta _{0}=(\rho ,\widetilde{\sigma }%
_{v}^{2},\zeta ^{\prime })^{\prime }$, by following the working paper
version of K2013.$\smallskip $

Let $g(y_{i}|y_{i,1},\theta _{0})$ denote the normal p.d.f. of $y_{i}$ given 
$y_{i,1},$ let $D$ denote a $(T-2)\times (T-1)$ first-difference matrix, let 
$\widetilde{y}_{i}=y_{i}-y_{i,1}\iota ,$ $\widetilde{y}%
_{i,-1}=y_{i,-1}-y_{i,1}\iota ,$ $w_{i}=\widetilde{y}_{i}-\rho \widetilde{y}%
_{i,-1}$ and $d=\Psi ^{-1}\iota /\iota ^{\prime }\Psi ^{-1}\iota .$ Note
that $g(y_{i}|y_{i,1},\theta _{0})=g(w_{i}|y_{i,1},\theta _{0}),$ $%
E(Dw_{i}w_{i}^{\prime }d|y_{i,1})=0$ and $d^{\prime }\iota =1.$ To derive $%
\frac{\partial l}{\partial \rho }$, it is useful to factorise $%
g(w_{i}|y_{i,1},\theta _{0})=g(Dw_{i}$ $d^{\prime }w_{i}|y_{i,1},\theta
_{0}) $ as 
\begin{eqnarray}
&&g(w_{i}|y_{i,1},\theta _{0})=g(Dw_{i}|y_{i,1},\rho ,\zeta )g(d^{\prime
}w_{i}|y_{i,1},\theta _{0})\text{ where } \\
&&\hspace{0.37in}\ln g(Dw_{i}|y_{i,1},\rho ,\zeta )=-\frac{(T-2)}{2}\ln
(2\pi )-\frac{1}{2}\ln \left\vert D\Psi D^{\prime }\right\vert -\qquad \\
&&\hspace{0.47in}\qquad \frac{1}{2}(\widetilde{y}_{i}-\rho \widetilde{y}%
_{i,-1})^{\prime }D^{\prime }(D\Psi D^{\prime })^{-1}D(\widetilde{y}%
_{i}-\rho \widetilde{y}_{i,-1}).  \notag
\end{eqnarray}%
Now let $\sigma _{u}^{2}=Var(d^{\prime }w_{i}|y_{i,1})=Var(d^{\prime
}u_{i})=1/\iota ^{\prime }\Psi ^{-1}\iota +\widetilde{\sigma }_{v}^{2}.$
Then we can parametrize\linebreak $\ln g(d^{\prime }w_{i}|y_{i,1},\theta
_{0})$ more parsimoniously as 
\begin{equation}
\ln g(d^{\prime }w_{i}|y_{i,1},\rho ,\sigma _{u}^{2})=-\frac{1}{2}\ln (2\pi
)-\frac{1}{2}\ln \sigma _{u}^{2}-\frac{1}{2\sigma _{u}^{2}}(d^{\prime }%
\widetilde{y}_{i}-\rho d^{\prime }\widetilde{y}_{i,-1})^{2}.
\end{equation}%
Finally, given that $l=l_{FE}(\theta _{0})=\sum_{i}\ln g(Dw_{i}|y_{i,1},\rho
,\zeta )+\sum_{i}\ln g(d^{\prime }w_{i}|y_{i,1},\theta _{0}),$ we obtain 
\begin{eqnarray}
\frac{\partial l}{\partial \rho } &=&\sum_{i}\widetilde{y}_{i,-1}^{\prime
}D^{\prime }(D\Psi D^{\prime })^{-1}D(\widetilde{y}_{i}-\rho \widetilde{y}%
_{i,-1})+\frac{1}{\sigma _{u}^{2}}\sum_{i}(d^{\prime }(\widetilde{y}%
_{i}-\rho \widetilde{y}_{i,-1})d^{\prime }\widetilde{y}_{i,-1}),\qquad
\label{sco} \\
\frac{\partial l}{\partial \zeta } &=&\frac{1}{2}(\frac{\partial vec(\Psi
_{D})}{\partial \zeta ^{\prime }})^{\prime }(\Psi _{D}^{-1}\otimes \Psi
_{D}^{-1})\sum_{i}(vec(Y_{i}-\Psi _{D}))-\frac{1}{\sigma _{u}^{2}}%
\sum_{i}(d^{\prime }u_{i})\frac{\partial (d^{\prime }u_{i})}{\partial \zeta }%
,\text{ and}  \notag \\
\frac{\partial l}{\partial \sigma _{u}^{2}} &=&-\frac{N}{2\sigma _{u}^{2}}+%
\frac{1}{2\sigma _{u}^{4}}\sum_{i}(d^{\prime }\widetilde{y}_{i}-\rho
d^{\prime }\widetilde{y}_{i,-1})^{2},\text{ where }  \notag \\
&&\quad \frac{\partial (d^{\prime }u_{i})}{\partial \zeta }=(-\iota ^{\prime
}\Psi ^{-1}\iota \Psi ^{-2}u_{i}+\Psi ^{-2}\iota \iota ^{\prime }\Psi
^{-1}u_{i})/(\iota ^{\prime }\Psi ^{-1}\iota )^{2}.  \notag
\end{eqnarray}

Using the reparametrization given above theorem 1, i.e., $r=r_{n},$ $%
s_{t}^{2}=rs_{t,n}^{2},$ $t=2,...,T$, so that $z=\rho z_{n},$ and $%
\widetilde{s}_{v}^{2}/s_{2}^{2}=\widetilde{s}_{v,n}^{2}+(1-r),$ we find that 
$\sigma _{u}^{2}=\sigma _{u}^{2}(\theta _{n})=r/(\iota ^{\prime }\Psi
_{n}^{-1}\iota )+s_{2,n}^{2}r(\widetilde{s}_{v,n}^{2}+(1-r))$ where $\Psi
_{n}=diag(s_{2,n}^{2},...,s_{T,n}^{2}).\smallskip $

Let $l_{n}=l_{n}(\theta _{n}).$ Then from $l_{n}=l(\theta (\theta _{n}))$
and $\frac{\partial l}{\partial \rho }$ and $\frac{\partial l}{\partial
\sigma _{u}^{2}}$ in (\ref{sco}) we obtain that 
\begin{eqnarray}
\frac{\partial l_{n}}{\partial r_{n}} &=&\frac{\partial l_{n}}{\partial r}=-%
\frac{1}{2r}N(T-2)+ \\
&&\frac{1}{2r^{2}}\sum_{i}(\widetilde{y}_{i}-r\widetilde{y}_{i,-1})^{\prime
}D^{\prime }(D\Psi _{n}D^{\prime })^{-1}D(\widetilde{y}_{i}-r\widetilde{y}%
_{i,-1})+  \notag \\
&&\frac{1}{r}\sum_{i}\widetilde{y}_{i,-1}^{\prime }D^{\prime }(D\Psi
_{n}D^{\prime })^{-1}D(\widetilde{y}_{i}-r\widetilde{y}_{i,-1})+  \notag \\
&&\frac{1}{\sigma _{u}^{2}}\sum_{i}(d^{\prime }(\widetilde{y}_{i}-r%
\widetilde{y}_{i,-1}))d^{\prime }\widetilde{y}_{i,-1}+\left( -\frac{N}{%
2\sigma _{u}^{2}}+\frac{1}{2\sigma _{u}^{4}}\sum_{i}(d^{\prime }(\widetilde{y%
}_{i}-r\widetilde{y}_{i,-1}))^{2}\right) \frac{\partial \sigma _{u}^{2}}{%
\partial r}  \notag
\end{eqnarray}%
where $\frac{\partial \sigma _{u}^{2}}{\partial r}=\frac{\partial \sigma
_{u}^{2}(\theta _{n})}{\partial r}.\smallskip $

We have assumed that $\mathcal{\theta }_{0}=\mathcal{\theta }_{\ast }$ and
TSH. Letting $\widetilde{\sigma }_{u}^{2}=r(1+(T-1)(1-r))$ and $\theta
_{0,n,N}=(r,\widetilde{\sigma }_{v,n}^{2},\zeta _{n}^{\prime })^{\prime
}=(a_{N},$ $0,$ $\sigma ^{2}\iota ^{\prime })^{\prime }$, replacing $%
s_{2,n}^{2},...,s_{T,n}^{2}$ and $\widetilde{s}_{v,n}^{2}$ by elements of $%
\widetilde{\theta }_{n},$ which are $\sqrt{N}$-consistent estimators, noting
that $D^{\prime }(DD^{\prime })^{-1}D=I_{T-1}-\iota \iota ^{\prime
}/(T-1)\equiv Q,$ and using a Taylor expansion of $\frac{\partial
l_{n}(\theta _{n})}{\partial r}|_{\widetilde{\theta }_{n}}$, we obtain%
\begin{equation*}
N^{-1/2}\frac{\partial l_{n}(\theta _{n})}{\partial r}|_{\widetilde{\theta }%
_{n}}\overset{asy}{=}N^{-1/2}\frac{\partial l_{n}(\theta _{n})}{\partial r}%
|_{\theta _{0,n,N}}+\overline{H}_{1,.}N^{1/2}(\widetilde{\theta }_{n}-\theta
_{0,n,N}),
\end{equation*}%
where $\overline{H}_{1,.}N^{1/2}(\widetilde{\theta }_{n}-\theta
_{0,n,N})=O_{p}(1)$ and 
\begin{eqnarray}
\frac{2\sigma ^{2}\widetilde{\sigma }_{u}^{4}}{N}\frac{\partial l_{n}(\theta
_{n})}{\partial r}|_{\theta _{0,n,N}} &=&-\sigma ^{2}\frac{\widetilde{\sigma 
}_{u}^{4}}{r}(T-2)+  \label{leq} \\
&&\frac{\widetilde{\sigma }_{u}^{4}}{r^{2}}\frac{1}{N}\sum_{i}(\widetilde{y}%
_{i}-r\widetilde{y}_{i,-1})^{\prime }Q(\widetilde{y}_{i}-r\widetilde{y}%
_{i,-1})+  \notag \\
&&2\frac{\widetilde{\sigma }_{u}^{4}}{r}\frac{1}{N}\sum_{i}\widetilde{y}%
_{i,-1}^{\prime }Q(\widetilde{y}_{i}-r\widetilde{y}_{i,-1})+  \notag \\
&&2\widetilde{\sigma }_{u}^{2}\frac{1}{(T-1)}\frac{1}{N}\sum_{i}(\iota
^{\prime }(\widetilde{y}_{i}-r\widetilde{y}_{i,-1}))\iota ^{\prime }%
\widetilde{y}_{i,-1}+  \notag \\
&&\left( -\sigma ^{2}\widetilde{\sigma }_{u}^{2}+\frac{1}{(T-1)}\frac{1}{N}%
\sum_{i}(\iota ^{\prime }(\widetilde{y}_{i}-r\widetilde{y}%
_{i,-1}))^{2}\right) \frac{\partial \widetilde{\sigma }_{u}^{2}}{\partial r}
\notag
\end{eqnarray}%
with $\frac{\partial \widetilde{\sigma }_{u}^{2}}{\partial r}%
=1+(T-1)(2(1-r)-1).$

We can write $\widetilde{y}_{i}-r\widetilde{y}_{i,-1}=\varepsilon _{i}+(1-r)%
\widetilde{y}_{i,-1}$ and $\widetilde{y}_{i,-1}=B\varepsilon _{i}$ where $B$
is a $(T-1)\times (T-1)$ matrix with $B_{i,j}=\rho ^{i-j-1}=1$ for $i>j$ and 
$B_{i,j}=0$ for $i\leq j.$ Let $\phi =1-r.$ From $\frac{\partial l_{n}}{%
\partial \rho _{n}}|_{\theta _{\ast }}=0$ (cf. Theorem 1) we deduce that the
RHS of (\ref{leq}) contains the common factor $\phi $ at least once. Note
that $tr(B^{\prime }QB)=\frac{1}{6}(T-2)T,$ $tr(QB)=-\frac{1}{2}(T-2),\
\iota ^{\prime }B^{\prime }B\iota =\frac{1}{6}(T-2)(T-1)(2T-3)$ and $\iota
^{\prime }B\iota =\frac{1}{2}(T-2)(T-1).$ Then we obtain%
\begin{equation}
E(\frac{2\widetilde{\sigma }_{u}^{4}}{N}\frac{\partial l_{n}}{\partial r}%
|_{\theta _{0,n,N}})=-\frac{1}{6}\phi ^{3}T\left( T-1\right) \left( \phi
(T-2)(T-1)-2T^{2}+2T-2\right)  \label{dlnr}
\end{equation}%
Moreover, 
\begin{multline*}
E(\frac{2\widetilde{\sigma }_{u}^{4}}{N}\frac{\partial l_{n}}{\partial r}%
|_{\theta _{0,n,N}})+\frac{\widetilde{\sigma }_{u}^{4}}{r}(T-2)+\widetilde{%
\sigma }_{u}^{2}\frac{\partial \widetilde{\sigma }_{u}^{2}}{\partial r}= \\
\frac{1}{6}\phi T\left( T-1\right) ((-\phi ^{2}\left( T-1\right) +2\phi
\left( T-2\right) +6)\phi \left( T-2\right) +6).
\end{multline*}%
Thus $\frac{1}{N}\frac{\partial l_{n}(\theta _{n})}{\partial r}|_{\theta
_{0,n,N}}$ contains the common factor $\phi $ exactly once and we can divide 
$\frac{2\widetilde{\sigma }_{u}^{4}}{N}\frac{\partial l_{n}}{\partial r}%
|_{\theta _{0,n,N}}$ by $\phi .$

Let $S$ be a $(T-1)\times (T-1)$ diagonal matrix with $S_{1,1}=\phi ^{-1}$
and $S_{i,i}=1$ for $i\geq 2$. Using reasoning similar to that used in
Davidson and MacKinnon (1993), cf. the proof of theorem 4, we obtain that 
\begin{equation*}
N^{-1/2}S\frac{\partial l_{n}(\theta _{n})}{\partial \theta _{n}}|_{%
\widetilde{\theta }_{n}}\overset{asy}{=}SA^{\prime }(A\overline{H}%
^{-1}A^{\prime })^{-1}A\overline{H}^{-1}N^{-1/2}\frac{\partial l_{n}(\theta
_{n})}{\partial \theta _{n}}|_{\theta _{0,n,N}}
\end{equation*}%
Note that all elements of the vector $N^{-1/2}S\frac{\partial l_{n}(\theta
_{n})}{\partial \theta _{n}}|_{\widetilde{\theta }_{n}}$ apart from its
first element are zero.

Using (\ref{dlnr}) and $\phi =1-r=\kappa /\sqrt[4]{N}$, we also have%
\begin{equation*}
\sqrt{N}E(\frac{\widetilde{\sigma }_{u}^{4}}{\phi N}\frac{\partial l_{n}}{%
\partial r}|_{\theta _{0,n,N}})=-\frac{1}{12}\kappa ^{2}T\left( T-1\right)
\left( (T-2)(T-1)\kappa /\sqrt[4]{N}-2T^{2}+2T-2\right)
\end{equation*}

Next we consider $E(\frac{\widetilde{\sigma }_{u}^{4}}{N}\frac{\partial l_{n}%
}{\partial \widetilde{s}_{v,n}^{2}}|_{\theta _{0,n,N}}).$%
\begin{multline*}
E(\frac{2\widetilde{\sigma }_{u}^{4}}{rN}\frac{\partial l_{n}}{\partial 
\widetilde{s}_{v,n}^{2}}|_{\theta _{0,n,N}})=E(\frac{2\widetilde{\sigma }%
_{u}^{4}}{rN}\frac{\partial l}{\partial \sigma _{u}^{2}}\frac{\partial
\sigma _{u}^{2}}{\partial \widetilde{s}_{v,n}^{2}}|_{\theta _{0,n,N}})= \\
E\left( -\widetilde{\sigma }_{u}^{2}(T-1)+\frac{1}{\sigma ^{2}}\frac{1}{N}%
\sum_{i}(\iota ^{\prime }(\widetilde{y}_{i}-r\widetilde{y}%
_{i,-1}))^{2}\right) = \\
(-(1-\phi )(1+(T-1)\phi )+1+\phi (T-2)+\frac{1}{6}\phi ^{2}(T-2)(2T-3))(T-1)=
\\
\frac{1}{6}T\phi ^{2}\left( 2T^{2}-3T+1\right)
\end{multline*}

Hence 
\begin{equation*}
\sqrt{N}E(\frac{\widetilde{\sigma }_{u}^{4}}{rN}\frac{\partial l_{n}}{%
\partial \widetilde{s}_{v,n}^{2}}|_{\theta _{0,n,N}})=\frac{1}{12}\kappa
^{2}T\left( 2T^{2}-3T+1\right) .
\end{equation*}%
Finally, we consider $E(\frac{\partial l_{n}}{\partial z_{n}}|_{\theta
_{0,n,N}}).$ From the working paper version of K2013 we have%
\begin{equation*}
\frac{\partial l_{n}}{\partial z_{n}}|_{\theta _{0,n,N}}=r(\frac{\partial l}{%
\partial z}|_{\theta _{0,n,N}})+(\frac{\partial l}{\partial \sigma _{u}^{2}}%
|_{\theta _{0,n,N}})\frac{\partial \sigma _{u}^{2}}{\partial z_{n}}
\end{equation*}

and 
\begin{equation*}
\frac{\partial \sigma _{u}^{2}}{\partial z_{n}}=\frac{r}{(T-1)^{2}}\iota
+(r(1-r),0,0,...,0)^{\prime }.
\end{equation*}

Note that 
\begin{eqnarray*}
\sigma ^{2}\frac{\partial l}{\partial z}|_{\theta _{0,n,N}} &=&\frac{1}{%
2r^{2}\sigma ^{2}}(\frac{\partial vec(\Psi )}{\partial \zeta ^{\prime }}%
)^{\prime }\sum_{i}vec(Q(\widetilde{y}_{i}-r\widetilde{y}_{i,-1})(\widetilde{%
y}_{i}-r\widetilde{y}_{i,-1})^{\prime }Q-r\sigma ^{2}Q) \\
&&+\frac{1}{r\sigma ^{2}\widetilde{\sigma }_{u}^{2}(T-1)}\sum_{i}(\iota
^{\prime }(\widetilde{y}_{i}-r\widetilde{y}_{i,-1})Q(\widetilde{y}_{i}-r%
\widetilde{y}_{i,-1}))
\end{eqnarray*}%
and 
\begin{equation*}
\frac{2\sigma ^{2}\widetilde{\sigma }_{u}^{4}}{N}\frac{\partial l}{\partial
\sigma _{u}^{2}}=-\widetilde{\sigma }_{u}^{2}(T-1)+\frac{1}{\sigma ^{2}}%
\frac{1}{N}\sum_{i}(\iota ^{\prime }(\widetilde{y}_{i}-r\widetilde{y}%
_{i,-1}))^{2}
\end{equation*}%
It follows that 
\begin{multline*}
E(\frac{\sigma ^{2}\widetilde{\sigma }_{u}^{4}}{rN}\frac{\partial l_{n}}{%
\partial z_{n}}|_{\theta _{0,n,N}})=\frac{\widetilde{\sigma }_{u}^{4}}{2r^{2}%
}(\frac{\partial vec(\Psi )}{\partial \zeta ^{\prime }})^{\prime }vec(\phi
Q+2\phi QBQ+\phi ^{2}QBB^{\prime }Q)+\frac{\widetilde{\sigma }_{u}^{2}}{%
r(T-1)}(2\phi QB\iota +\phi ^{2}QBB^{\prime }\iota ) \\
+\frac{1}{2}(-\widetilde{\sigma }_{u}^{2}(T-1)+(\iota ^{\prime }\iota +2\phi
\iota ^{\prime }B\iota +\phi ^{2}\iota ^{\prime }BB^{\prime }\iota ))(\frac{1%
}{(T-1)^{2}}\iota +(\phi ,0,0,...,0)^{\prime })
\end{multline*}%
Simplifying yields%
\begin{multline*}
E(\frac{\sigma ^{2}\widetilde{\sigma }_{u}^{4}}{rN}\frac{\partial l_{n}}{%
\partial s_{2,n}^{2}}|_{\theta _{0,n,N}})= \\
\frac{1}{12}\phi ^{2}\left( 2T^{3}\phi ^{2}+2T^{3}\phi -9T^{2}\phi
^{2}+9T^{2}\phi +13T\phi ^{2}-35T\phi +12T-6\phi ^{2}+24\phi -18\right)
\end{multline*}

and for $k=2,...,T-1,$ 
\begin{multline*}
E(\frac{\sigma ^{2}\widetilde{\sigma }_{u}^{4}}{rN}\frac{\partial l_{n}}{%
\partial s_{k+1,n}^{2}}|_{\theta _{0,n,N}})= \\
\frac{1}{12}\phi ^{2}\left( 
\begin{array}{c}
2T^{3}\phi ^{2}-6T^{2}k\phi ^{2}-3T^{2}\phi ^{2}+12T^{2}\phi +6Tk^{2}\phi
^{2}+6Tk\phi ^{2}-24Tk\phi + \\ 
+T\phi ^{2}-12T\phi +12T-6k^{2}\phi ^{2}+6k^{2}\phi +18k\phi -18k%
\end{array}%
\right)
\end{multline*}

Hence $\sqrt{N}E(\frac{\sigma ^{2}\widetilde{\sigma }_{u}^{4}}{rN}\frac{%
\partial l_{n}}{\partial s_{2,n}^{2}}|_{\theta _{0,n,N}})=\frac{1}{12}\kappa
^{2}\left( 12T-18\right) +O(N^{-1/4})$ and for $k=2,...,T-1,$\linebreak $%
\sqrt{N}E(\frac{\sigma ^{2}\widetilde{\sigma }_{u}^{4}}{rN}\frac{\partial
l_{n}}{\partial s_{k+1,n}^{2}}|_{\theta _{0,n,N}})=\frac{1}{12}\kappa
^{2}\left( 12T-18k\right) +O(N^{-1/4}).$

Let $c_{1}$ be a ($T-1$)-vector such that $c_{1}\kappa
^{2}=\lim_{N\rightarrow \infty }N^{-1/2}E(S\frac{\partial l_{n}(\theta _{n})%
}{\partial \theta _{n}}|_{\theta _{0,n,N}}).$

Then we have $SN^{-1/2}\frac{\partial l_{n}(\theta _{n})}{\partial \theta
_{n}}|_{\theta _{0,n,N}}\overset{d}{\rightarrow }N(c_{1}\kappa ^{2},$%
\thinspace p$\lim_{N\rightarrow \infty }S\overline{\mathcal{J}}^{c}S)$ and $%
QLM_{FE}^{c}(\widetilde{\mathcal{\theta }}_{n})\overset{d}{\rightarrow }$%
\linebreak $\chi ^{2}($p$\lim_{N\rightarrow \infty }c_{1}^{\prime }S^{-1}%
\overline{H}^{-1}A^{\prime }(A\overline{H}^{-1}\overline{\mathcal{J}}^{c}%
\overline{H}^{-1}A^{\prime })^{-1}A\overline{H}^{-1}S^{-1}c_{1}\kappa
^{4},1).$\quad $\square \medskip $

\textbf{\noindent Proof of theorem 6:}

Note that 
\begin{eqnarray}
\frac{2\sigma ^{2}\widetilde{\sigma }_{u}^{4}}{rN}\frac{\partial l_{n}}{%
\partial s_{n}^{2}}|_{\theta _{0,n,N}} &=&-(T-2)\frac{\widetilde{\sigma }%
_{u}^{4}}{r}-\frac{\widetilde{\sigma }_{u}^{4}}{r}+ \\
&&\frac{\widetilde{\sigma }_{u}^{4}}{r^{2}\sigma ^{2}}\frac{1}{N}\sum_{i}(%
\widetilde{y}_{i}-r\widetilde{y}_{i,-1})^{\prime }Q(\widetilde{y}_{i}-r%
\widetilde{y}_{i,-1})+  \notag \\
&&\frac{\widetilde{\sigma }_{u}^{2}}{r\sigma ^{2}}\frac{1}{(T-1)}\frac{1}{N}%
\sum_{i}(\iota ^{\prime }(\widetilde{y}_{i}-r\widetilde{y}_{i,-1}))^{2}. 
\notag
\end{eqnarray}%
Using results in the proof of Theorem 5, it follows that%
\begin{equation*}
E(\frac{1}{N}\frac{\partial l_{n}}{\partial s_{n}^{2}}|_{\theta _{0,n,N}})=%
\frac{1}{12}\frac{\phi ^{2}T\left( T-1\right) \left( T^{2}\phi ^{2}-3T\phi
^{2}+4T\phi +2\phi ^{2}-5\phi +3\right) }{\sigma ^{2}(1-\phi )(1+(T-1)\phi
)^{2}}.
\end{equation*}%
Similarly to the proof of Theorem 5, we also have 
\begin{equation*}
E(\frac{1}{N}\frac{\partial l_{n}}{\partial r}|_{\theta _{0,n,N}})=-\frac{1}{%
12}\frac{\phi ^{3}T\left( T-1\right) \left( 2T+2\phi -3T\phi +T^{2}\phi
-2T^{2}-2\right) }{((1-\phi )(1+(T-1)\phi ))^{2}}
\end{equation*}%
and 
\begin{equation*}
E(\frac{1}{N}\frac{\partial l_{n}}{\partial \widetilde{s}_{v,n}^{2}}%
|_{\theta _{0,n,N}})=\frac{1}{12}\frac{\phi ^{2}T(2T^{2}-3T+1)}{(1-\phi
)(1+(T-1)\phi )^{2}}.
\end{equation*}

Let $\widetilde{S}$ be a $3\times 3$ diagonal matrix with $\widetilde{S}%
_{1,1}=\phi ^{-1},$ $\widetilde{S}_{2,2}=1$ and $\widetilde{S}_{3,3}=\sigma
^{2}.$ Then$\medskip $ we\linebreak obtain $c_{3}\equiv \lim_{N\rightarrow
\infty }N^{-1/2}E(\widetilde{S}\frac{\partial l_{n}(\theta _{n})}{\partial
\theta _{n}}|_{\theta _{0,n,N}})=\left[ 
\begin{array}{c}
\frac{1}{6}T\left( T-1\right) \left( T^{2}-T+1\right) \\ 
\frac{1}{12}T\left( 2T-1\right) \left( T-1\right) \\ 
\frac{1}{4}T\left( T-1\right)%
\end{array}%
\right] \kappa ^{2}.$ We also$\medskip $\linebreak have p$\lim_{N\rightarrow
\infty }\widetilde{S}\overline{H}\widetilde{S}=\left[ 
\begin{array}{ccc}
\frac{1}{2}T\left( T-1\right) \left( -T+T^{2}+1\right) & \frac{1}{6}T\left(
2T^{2}-3T+1\right) & \frac{1}{2}T\left( T-1\right) \\ 
\frac{1}{6}T\left( 2T^{2}-3T+1\right) & \frac{1}{2}(T-1)^{2} & \frac{1}{2}%
(T-1) \\ 
\frac{1}{2}T\left( T-1\right) & \frac{1}{2}(T-1) & \frac{1}{2}(T-1)%
\end{array}%
\right] .\medskip $

Using $tr(B^{\prime }B)=\frac{1}{2}(T-2)(T-1),$ $tr(B^{\prime }BB^{\prime
}B)=\frac{1}{6}(T-2)(T-1)(T^{2}-3T+3),$ $tr(QBQB)=-\frac{1}{12}(T-2)(T-6)$
and $tr(B^{\prime }BQB)=-\frac{1}{24}(T-2)T(T+1),$ we obtain$\medskip $

p$\lim_{N\rightarrow \infty }\widetilde{S}\overline{\mathcal{J}}^{c}%
\widetilde{S}=\left[ 
\begin{array}{ccc}
\frac{1}{3}T\left( T-1\right) \left( -T+T^{2}+1\right) & \frac{1}{6}T\left(
2T^{2}-3T+1\right) & \frac{1}{2}T\left( T-1\right) \\ 
\frac{1}{6}T\left( 2T^{2}-3T+1\right) & \frac{1}{2}(T-1)^{2} & \frac{1}{2}%
(T-1) \\ 
\frac{1}{2}T\left( T-1\right) & \frac{1}{2}(T-1) & \frac{1}{2}(T-1)%
\end{array}%
\right] .\medskip $ Finally, we have $\widetilde{S}N^{-1/2}\frac{\partial
l_{n}(\theta _{n})}{\partial \theta _{n}}|_{\theta _{0,n,N}}\overset{d}{%
\rightarrow }N(c_{3},$ p$\lim_{N\rightarrow \infty }\widetilde{S}\overline{%
\mathcal{J}}^{c}\widetilde{S})$ and $QLM_{FE}^{c}(\widetilde{\mathcal{\theta 
}}_{n})\overset{d}{\rightarrow }\chi ^{2}(\gamma ,1)$ with $\gamma =$ p$%
\lim_{N\rightarrow \infty }c_{3}^{\prime }\widetilde{S}^{-1}\overline{H}%
^{-1}A^{\prime }(A\overline{H}^{-1}\overline{\mathcal{J}}^{c}\overline{H}%
^{-1}A^{\prime })^{-1}A\overline{H}^{-1}\widetilde{S}^{-1}c_{3}=\frac{%
(2T-3)T(T-1)(T-2)}{72}\kappa ^{4}.$\ $\square \medskip $

To prove theorem 7 we will make use of some intermediate results. It will
also be useful to introduce some additional notation. For $n\in {\mathbb{N}}$
and $a,b,c\in {\mathbb{R}}$, $\Delta _{n}(a,b,c)$ denotes the $(n\times n)$
tridiagonal matrix of the following form 
\begin{equation*}
\Delta _{n}(a,b,c)=%
\begin{pmatrix}
a & c & 0 & 0 & \dots & 0 & 0 & 0 \\ 
c & b & c & 0 & \dots & 0 & 0 & 0 \\ 
0 & c & b & c & \dots & 0 & 0 & 0 \\ 
\vdots & \vdots & \vdots & \vdots & \vdots & \vdots & \vdots & \vdots \\ 
0 & 0 & 0 & 0 & \dots & c & b & c \\ 
0 & 0 & 0 & 0 & \dots & 0 & c & b%
\end{pmatrix}%
.
\end{equation*}

\begin{lemma}
Let $T\geq 2$ and $G_{T}=\Delta _{T}(1,2,-1)$. Then the inverse $G_{T}^{-1}$
has the following simple structure: 
\begin{equation*}
G_{T}^{-1}=%
\begin{pmatrix}
T & T-1 & T-2 & \dots & 2 & 1 \\ 
T-1 & T-1 & T-2 & \dots & 2 & 1 \\ 
T-2 & T-2 & T-2 & \dots & 2 & 1 \\ 
\vdots & \vdots & \vdots & \vdots & \vdots & \vdots \\ 
2 & 2 & 2 & 2 & 2 & 1 \\ 
1 & 1 & 1 & 1 & 1 & 1%
\end{pmatrix}%
.
\end{equation*}
\end{lemma}

\textbf{\noindent Proof of lemma 1: }It is straightforward to verify that $%
G_{T}G_{T}^{-1}=I_{T}$.\quad $\square \medskip $

Let $D_{u,T}$ be the $(T^{2}\times \frac{T(T+1)}{2})$ matrix such that $%
\mathrm{vec}(A)=D_{u,T}\mathrm{vech}(A)$ for any $(T\times T)$ matrix $A.$ $%
D_{u,T}$ is a so-called duplication matrix as defined in Ch. 3 of Magnus and
Neudecker (2007). Let $D_{u,T}^{+}=(D_{u,T}^{\prime
}D_{u,T})^{-1}D_{u,T}^{\prime }$ be the Penrose-Moore inverse of $D_{u,T}.$

\begin{proposition}
Let $T\geq 2$ and $M_{T}=2D_{u,T}^{+}(G_{T}\otimes
G_{T})(D_{u,T}^{+})^{\prime }$. Then $M_{T}$ is invertible and its inverse $%
M_{T}^{-1}$ has the form 
\begin{equation}
M_{T}^{-1}=\frac{1}{2}D_{u,T}^{\prime }(G_{T}^{-1}\otimes G_{T}^{-1})D_{u,T}.
\label{prop1eq}
\end{equation}
\end{proposition}

\textbf{\noindent Proof of proposition 1: }This follows from application of
equation (13) in Theorem 13 in Ch. 3 of Magnus and Neudecker (2007).\quad $%
\square \medskip $

\begin{lemma}
Let $T\geq 2$, $G_{T}=\Delta _{T}(1,2,-1),$ $H_{T}=\Delta _{T}(0,-2,1),$%
\begin{equation*}
A_{T}=\Delta _{T}(-\frac{T-1}{2},-\frac{T-2}{3},\frac{T+1}{6}),\text{ and}
\end{equation*}%
\begin{equation*}
F_{T}=(T,T-1,T-2,...,2,1)^{\prime }\otimes e_{1}^{T},
\end{equation*}%
where $e_{1}$ is the first standard column basis vector of ${\mathbb{R}}%
^{T}. $

Then we have $G_{T}^{-1}H_{T}=F_{T}-I_{T}$ and $%
-6G_{T}^{-1}A_{T}=2(T+1)F_{T}-6G_{T}^{-1}+(T+1)I_{T}.$
\end{lemma}

\textbf{\noindent Proof of lemma 2: }Verification of both claims is
straightforward.\quad $\square \medskip $

Let $G_{T}=\Delta _{T}(1,2,-1).$ Let 
\begin{equation}
P_{T}=(0\text{\ }g_{T}\text{\ }I_{p_{T}})  \label{bm0}
\end{equation}%
be a $p_{T}\times \frac{1}{2}T(T+1)$ matrix where $p_{T}=\frac{1}{2}T(T+1)-2$
and $g_{T}$ is a $p_{T}$-vector such that 
\begin{equation}
(1,-1,g_{T}^{\prime })^{\prime }=\mathrm{vech}(G_{T})  \label{bm2}
\end{equation}%
and let 
\begin{equation}
\overline{P}_{T}=%
\begin{pmatrix}
Z_{T}M_{T}^{-1} \\ 
P_{T}%
\end{pmatrix}
\label{bm1}
\end{equation}%
where $M_{T}^{-1}$ is given in (\ref{prop1eq}) and$\vspace{-0.12in}$%
\begin{equation}
Z_{T}=%
\begin{pmatrix}
1 & 0 & 0_{1,p_{T}} \\ 
0 & 1 & -g_{T}^{\prime }%
\end{pmatrix}
\label{zm1}
\end{equation}

\begin{lemma}
Let $T\geq 2,$ let $A_{T}=\Delta _{T}(-\frac{T-1}{2},-\frac{T-2}{3},\frac{T+1%
}{6})$ and let $\overline{P}_{T}$ be defined by (\ref{bm1}), (\ref{zm1}), (%
\ref{bm2}), (\ref{bm0}) and (\ref{prop1eq}). Then we have$\vspace{-0.1in}$ 
\begin{equation}
\overline{P}_{T}\mathrm{vech}(A_{T})=\mathrm{vech}(I_{T})-e_{1},\vspace{%
-0.1in}  \label{lemma3eq}
\end{equation}%
where $e_{1}$ is the first standard column basis vector of ${\mathbb{R}}%
^{T(T+1)/2}$.
\end{lemma}

\textbf{\noindent Proof of lemma 3: }

First note that $P_{T}\mathrm{vech}(A_{T})=\frac{T+1}{6}g_{T}+(0_{p_{T},2}$ $%
I_{p_{T}})\mathrm{vech}(A_{T})=(0_{p_{T},2}$ $I_{p_{T}})\mathrm{vech}%
(I_{T}). $

Next we need to show that $%
\begin{pmatrix}
1 & 0 & 0_{1,p_{T}}%
\end{pmatrix}%
M_{T}^{-1}\mathrm{vech}(A_{T})=0.$ For this it suffices to show that 
\begin{equation}
((T,T-1,T-2,...,2,1)^{\prime }\otimes (T,T-1,T-2,...,2,1)^{\prime })\mathrm{%
vec}(A_{T})=0.  \label{lm3eq}
\end{equation}%
The LHS of (\ref{lm3eq}) equals $(-\frac{T-2}{3})\frac{1}{6}T(T+1)(2T+1)+%
\frac{T-2}{3}T^{2}-\frac{T-1}{2}T^{2}+\frac{T+1}{6}%
(2T(T-1)+2(T-1)(T-2)+...+2\cdot 2\cdot 1)=-\frac{1}{18}(T-2)T(T+1)(2T+1)-%
\frac{T^{2}}{6}(T+1)+\frac{T+1}{3}\left(
\sum_{t=1}^{T-1}t^{2}+\sum_{t=1}^{T-1}t\right) =-\frac{1}{18}%
(T-2)T(T+1)(2T+1)-\frac{T^{2}}{6}(T+1)+\frac{T+1}{3}(\frac{1}{6}(T-1)T(2T-1)+%
\frac{1}{2}(T-1)T)=0.$

Finally, we need to show that $%
\begin{pmatrix}
0 & 1 & -g_{T}^{\prime }%
\end{pmatrix}%
M_{T}^{-1}\mathrm{vech}(A_{T})=0.$ Note that $%
\begin{pmatrix}
0 & 1 & -g_{T}^{\prime }%
\end{pmatrix}%
^{\prime }=\mathrm{vech}(H_{T})$ where $H_{T}=\Delta _{T}(0,-2,1)$ and that $%
(\mathrm{vech}(H_{T}))^{\prime }M_{T}^{-1}\mathrm{vech}(A_{T})=\frac{1}{2}(%
\mathrm{vec}(H_{T}))^{\prime }\times $ $(G_{T}^{-1}\otimes G_{T}^{-1})%
\mathrm{vec}(A_{T})=\frac{1}{2}tr(G_{T}^{-1}A_{T}G_{T}^{-1}H_{T}).$ The last
equality follows from application of Theorem 3 in Ch. 2 of Magnus and
Neudecker (2007). From lemma 2 and additivity of the trace, it follows
straightforwardly that $\frac{1}{2}tr(G_{T}^{-1}A_{T}G_{T}^{-1}H_{T})=0.$%
\quad $\square \medskip $

\begin{lemma}
Let $T\geq 2,$ $G_{T}=\Delta _{T}(1,2,-1)$ and $A_{T}=\Delta _{T}(-\frac{T-1%
}{2},-\frac{T-2}{3},\frac{T+1}{6}).$ Then we have 
\begin{equation}
\mathrm{vec}(A_{T})^{\top }(G_{T}^{-1}\otimes G_{T}^{-1})\mathrm{vec}(A_{T})=%
\frac{(2T-1)(T+1)T(T-1)}{36}.  \label{lm4eq}
\end{equation}
\end{lemma}

\textbf{\noindent Proof of lemma 4: }The claim in (\ref{lm4eq}) follows
straightforwardly from $(\mathrm{vec}(A_{T}))^{\prime }(G_{T}^{-1}\otimes
G_{T}^{-1})\mathrm{vec}(A_{T})=tr(G_{T}^{-1}A_{T}G_{T}^{-1}A_{T})$, lemma 2
and additivity of the trace.\quad $\square \medskip $

\begin{lemma}
Let $T\geq 2$, $G_{T}=\Delta _{T}(1,2,-1)$ and $M_{T}=2D_{u,T}^{+}(G_{T}%
\otimes G_{T})(D_{u,T}^{+})^{\prime }.$ Let $P_{T}$ be defined by (\ref{bm0}%
) and (\ref{bm2}) and let $R_{T}=P_{T}M_{T}P_{T}^{\prime }$. Let $w_{T}$ be
the column vector of size $\frac{T(T+1)}{2}-2$ defined by $%
(1,0,w_{T}^{\prime })^{\prime }=\mathrm{vech}(I_{T}).$ Then $R_{T}$ is
invertible and we have 
\begin{equation}
w_{T}^{\prime }R_{T}^{-1}w_{T}=\frac{(2T-1)(T+1)T(T-1)}{72}.
\end{equation}
\end{lemma}

\textbf{\noindent Proof of lemma 5: }First note that%
\begin{equation}
\overline{P}_{T}M_{T}\overline{P}_{T}^{\prime }=%
\begin{pmatrix}
M_{0,T} & 0_{2,p_{T}} \\ 
0_{p_{T},2} & R_{T}%
\end{pmatrix}%
,  \label{lm5eq}
\end{equation}%
where $\overline{P}_{T}$ is defined by (\ref{bm1}) and (\ref{zm1}), $M_{0,T}$
is an $(2\times 2)$ matrix and $0_{p,q}$ denotes the zero matrix of size $%
(p\times q)$. From (\ref{lm5eq}) and invertibility of $\overline{P}_{T}$ and 
$M_{T}$, we obtain 
\begin{equation*}
(\overline{P}_{T}^{\prime })^{-1}M_{T}^{-1}(\overline{P}_{T})^{-1}=%
\begin{pmatrix}
M_{0,T}^{-1} & 0_{2,p_{T}} \\ 
0_{p_{T},2} & R_{T}^{-1}%
\end{pmatrix}%
,
\end{equation*}%
and if we choose $\widehat{w}_{T}=(0,0,w_{T}^{\prime })^{\prime },$ we
obtain 
\begin{equation*}
\widehat{w}_{T}^{\prime }(\overline{P}_{T}^{\prime })^{-1}M_{T}^{-1}(%
\overline{P}_{T})^{-1}\widehat{w}_{T}=w_{T}^{\prime }R_{T}^{-1}w_{T},
\end{equation*}%
which we are after. Next, by using the formula for $M_{T}^{-1}$ given in
Proposition 1, we obtain%
\begin{equation}
w_{T}^{\prime }R_{T}^{-1}w_{T}=\frac{1}{2}z_{T}^{\prime }(G_{T}^{-1}\otimes
G_{T}^{-1})z_{T}  \label{lm5eq2}
\end{equation}%
with $z_{T}=D_{u,T}(\overline{P}_{T})^{-1}\widehat{w}_{T}$.

Lemma 3 implies $\mathrm{vech}(A_{T})=(\overline{P}_{T})^{-1}\widehat{w}_{T}$
where $A_{T}=\Delta _{T}(-\frac{T-1}{2},-\frac{T-2}{3},\frac{T+1}{6})$ and
hence $\mathrm{vec}(A_{T})=D_{u,T}(\overline{P}_{T})^{-1}\widehat{w}%
_{T}=z_{T}.$ From this, equation (\ref{lm5eq2}) and lemma 4, it follows that 
$w_{T}^{\prime }R_{T}^{-1}w_{T}=\frac{1}{2}\mathrm{vec}(A_{T})^{\top
}(G_{T}^{-1}\otimes G_{T}^{-1})\mathrm{vec}(A_{T})=\frac{(2T-1)(T+1)T(T-1)}{%
72}$.\quad $\square \medskip $

\textbf{\noindent Proof of theorem 7:}

When the data are i.i.d. and normal, $\theta _{0}=\theta _{\ast }$ and when $%
r=1-\kappa /\sqrt[4]{N}$,\linebreak $Evech\left( D_{r}(\Delta y_{i}(\Delta
y_{i})^{\prime })D_{r}^{\prime }\right) =\sigma ^{2}vech\left(
D_{r}D_{r}^{\prime }\right) $ and $\lim_{N\rightarrow \infty
}N^{1/2}E(m(r))=\sigma ^{2}\kappa ^{2}\times \linebreak vech\left(
diag(0,1,1,...,1,1\right) )\equiv \sigma ^{2}\kappa ^{2}c_{1}.$

Let $G=D_{1}D_{1}^{\prime }.$ We have $E(vech(D_{1}\varepsilon
_{i}\varepsilon _{i}^{\prime }D_{1}^{\prime }-\sigma
^{2}G)vech(D_{1}\varepsilon _{i}\varepsilon _{i}^{\prime }D_{1}^{\prime
}-\sigma ^{2}G)^{\prime })=2\sigma ^{4}D_{u}^{+}(G\otimes
G)(D_{u}^{+})^{\prime }$ with $D_{u}^{+}=(D_{u}^{\prime
}D_{u})^{-1}D_{u}^{\prime }$ where $D_{u}$ is the so-called duplication
matrix such that $vec(A)=D_{u}vech(A)$ for any $(T-1)\times (T-1)$ matrix $%
A, $ see Magnus and Neudecker (2007). It follows that p$\lim_{N\rightarrow
\infty }\widehat{V}_{mm}(\rho )=2\sigma ^{4}PD_{u}^{+}(G\otimes
G)(D_{u}^{+})^{\prime }P^{\prime }$ and $GMM-AR(\rho )\overset{d}{%
\rightarrow }\chi ^{2}(\gamma ,p)$ where $\gamma =c_{1}^{\prime
}(2PD_{u}^{+}(G\otimes G)(D_{u}^{+})^{\prime }P^{\prime })^{-1}c_{1}\kappa
^{4}$ and $p=\dim (m(r))=\frac{1}{2}T(T-1)-2.$

Finally, we show that $\gamma =\frac{(2T-3)T(T-1)(T-2)}{72}\kappa ^{4}.$
This follows immediately from\linebreak lemma 5 upon replacing $T$ by $T-1$
(Note $G=G_{T-1},$ $D_{u}^{+}=D_{u,T-1}^{+},$ $P=P_{T-1}$ and $c_{1}=w_{T-1}$%
)$.$\quad $\square \medskip $\pagebreak

\textbf{\noindent Proof of theorem 8:}

Let $w$ be a $p$-vector with $w^{\prime }w=1.$ When the data are i.i.d. and
normal, and $\theta _{0}=\theta _{\ast }$ (so that TSH\ holds), the large
sample distribution of the GMM-AR test statistic that uses $w^{\prime
}m(\rho )$ for testing $H_{0}:$ $\rho =1-\kappa /\sqrt[4]{N}$ is given by $%
\chi ^{2}(\gamma _{w},1)$ where $\gamma _{w}=c_{1}^{\prime }w(2w^{\prime
}PD_{u}^{+}(G\otimes G)(D_{u}^{+})^{\prime }P^{\prime }w)^{-1}w^{\prime
}c_{1}\kappa ^{4}$ with $c_{1}=vech\left( diag(0,1,1,...,1,1\right) )$ (cf.
the proof of Theorem 7)$.$ The combination $w$ that maximizes the local
power of the GMM-AR test is the one that maximizes the value of $\gamma
_{w}. $

The maximal value of $\gamma _{w}$ is the largest root of the following
eigenvalue problem%
\begin{equation*}
\left\vert 2\lambda PD_{u}^{+}(G\otimes G)(D_{u}^{+})^{\prime }P^{\prime
}-\kappa ^{4}c_{1}c_{1}^{\prime }\right\vert =0
\end{equation*}%
and the power maximizing combination $w$ is given by the eigenvector
associated with this largest root. The largest root is equal to $\gamma
=c_{1}^{\prime }(2PD_{u}^{+}(G\otimes G)(D_{u}^{+})^{\prime }P^{\prime
})^{-1}c_{1}=\frac{(2T-3)T(T-1)(T-2)}{72}.$\quad $\square \vspace*{-0.1in}$

\begin{table}[hbp] \centering%
\caption{Empirical size of Quasi LM test based on RE Likelihood; Nominal size is 0.05; T=4; 2500
replications.\label{key}}%
\begin{tabular}{||l|c|c|c|c|c|c||}
\hline\hline
model & \multicolumn{2}{|c|}{S-Normal} & \multicolumn{2}{|c|}{S-$\chi ^{2}$}
& \multicolumn{2}{|c||}{NS-Normal} \\ \hline
$\rho $ & $N=100$ & $N=250$ & $N=100$ & $N=250$ & $N=100$ & $N=250$ \\ \hline
0.20 & .0532 & .0504 & .0552 & .0500 & .0508 & .0520 \\ 
0.50 & .0504 & .0468 & .0488 & .0540 & .0492 & .0580 \\ 
0.80 & .0452 & .0448 & .0540 & .0488 & .0520 & .0448 \\ 
0.90 & .0528 & .0492 & .0452 & .0492 & .0568 & .0548 \\ 
0.95 & .0520 & .0556 & .0460 & .0476 & .0488 & .0628 \\ 
0.98 & .0552 & .0444 & .0392 & .0452 & .0512 & .0536 \\ 
0.99 & .0504 & .0468 & .0400 & .0440 & .0380 & .0508 \\ \hline\hline
\end{tabular}
\bigskip 
\caption{Empirical size of Quasi LM test based on FE Likelihood; Nominal size is 0.05; T=4; 2500
replications.\label{key}}%
\begin{tabular}{||l|c|c|c|c|c|c||}
\hline\hline
model & \multicolumn{2}{|c|}{S-Normal} & \multicolumn{2}{|c|}{S-$\chi ^{2}$}
& \multicolumn{2}{|c||}{NS-Normal} \\ \hline
$\rho $ & $N=100$ & $N=250$ & $N=100$ & $N=250$ & $N=100$ & $N=250$ \\ \hline
0.20 & .0500 & .0488 & .0568 & .0484 & .0512 & .0536 \\ 
0.50 & .0468 & .0464 & .0512 & .0512 & .0468 & .0580 \\ 
0.80 & .0464 & .0464 & .0484 & .0496 & .0488 & .0480 \\ 
0.90 & .0568 & .0488 & .0432 & .0468 & .0544 & .0584 \\ 
0.95 & .0496 & .0500 & .0488 & .0472 & .0484 & .0540 \\ 
0.98 & .0508 & .0476 & .0464 & .0476 & .0528 & .0540 \\ 
0.99 & .0480 & .0452 & .0484 & .0504 & .0480 & .0516 \\ \hline\hline
\end{tabular}
\bigskip 
\caption{Empirical size of Quasi LM test based on RE Likelihood; Nominal size is 0.05; T=9; 2500
replications.\label{key}}%
\begin{tabular}{||l|c|c|c|c|c|c||}
\hline\hline
model & \multicolumn{2}{|c|}{S-Normal} & \multicolumn{2}{|c|}{S-$\chi ^{2}$}
& \multicolumn{2}{|c||}{NS-Normal} \\ \hline
$\rho $ & $N=100$ & $N=250$ & $N=100$ & $N=250$ & $N=100$ & $N=250$ \\ \hline
0.20 & .0556 & .0588 & .0488 & .0580 & .0432 & .0512 \\ 
0.50 & .0528 & .0508 & .0488 & .0484 & .0564 & .0456 \\ 
0.80 & .0536 & .0484 & .0532 & .0484 & .0576 & .0476 \\ 
0.90 & .0572 & .0480 & .0500 & .0548 & .0476 & .0468 \\ 
0.95 & .0548 & .0484 & .0504 & .0536 & .0492 & .0568 \\ 
0.98 & .0520 & .0428 & .0484 & .0380 & .0580 & .0560 \\ 
0.99 & .0516 & .0464 & .0448 & .0424 & .0480 & .0516 \\ \hline\hline
\end{tabular}
\end{table}%

\begin{table}[hbp] \centering%
\caption{Empirical size of Quasi LM test based on FE Likelihood; Nominal size is 0.05; T=9; 2500
replications.\label{key}}%
\begin{tabular}{||l|c|c|c|c|c|c||}
\hline\hline
model & \multicolumn{2}{|c|}{S-Normal} & \multicolumn{2}{|c|}{S-$\chi ^{2}$}
& \multicolumn{2}{|c||}{NS-Normal} \\ \hline
$\rho $ & $N=100$ & $N=250$ & $N=100$ & $N=250$ & $N=100$ & $N=250$ \\ \hline
0.20 & .0520 & .0576 & .0484 & .0588 & .0432 & .0508 \\ 
0.50 & .0500 & .0532 & .0496 & .0464 & .0572 & .0444 \\ 
0.80 & .0524 & .0496 & .0516 & .0556 & .0576 & .0456 \\ 
0.90 & .0576 & .0504 & .0524 & .0572 & .0492 & .0464 \\ 
0.95 & .0592 & .0496 & .0516 & .0564 & .0480 & .0524 \\ 
0.98 & .0596 & .0436 & .0536 & .0468 & .0608 & .0568 \\ 
0.99 & .0520 & .0456 & .0524 & .0532 & .0460 & .0524 \\ \hline\hline
\end{tabular}
\end{table}%

\begin{table}[hbp] \centering%
\caption{Empirical power of Quasi LM test based on RE Likelihood; $H_{0}:\rho=0.8$; Nominal size is 0.05; T=4; 2500
replications.\label{key}}%
\begin{tabular}{||l|c|c|c|c|c|c||}
\hline\hline
model & \multicolumn{2}{|c|}{S-Normal} & \multicolumn{2}{|c|}{S-$\chi ^{2}$}
& \multicolumn{2}{|c||}{NS-Normal} \\ \hline
true $\rho $ & $N=100$ & $N=250$ & $N=100$ & $N=250$ & $N=100$ & $N=250$ \\ 
\hline
0.50 & 0.357 & 0.567 & 0.381 & 0.556 & 0.876 & 0.977 \\ 
0.60 & 0.210 & 0.374 & 0.221 & 0.396 & 0.445 & 0.484 \\ 
0.70 & 0.090 & 0.154 & 0.107 & 0.174 & 0.092 & 0.104 \\ 
0.90 & 0.050 & 0.079 & 0.067 & 0.090 & 0.048 & 0.067 \\ 
0.95 & 0.069 & 0.121 & 0.070 & 0.125 & 0.066 & 0.101 \\ 
0.99 & 0.093 & 0.200 & 0.080 & 0.123 & 0.094 & 0.189 \\ \hline\hline
\end{tabular}
\end{table}%

\begin{table}[hbp] \centering%
\caption{Empirical power of Quasi LM test based on FE Likelihood; $H_{0}:\rho=0.8$; Nominal size is 0.05; T=4; 2500
replications.\label{key}}%
\begin{tabular}{||l|c|c|c|c|c|c||}
\hline\hline
model & \multicolumn{2}{|c|}{S-Normal} & \multicolumn{2}{|c|}{S-$\chi ^{2}$}
& \multicolumn{2}{|c||}{NS-Normal} \\ \hline
true $\rho $ & $N=100$ & $N=250$ & $N=100$ & $N=250$ & $N=100$ & $N=250$ \\ 
\hline
0.50 & 0.191 & 0.339 & 0.231 & 0.398 & 0.865 & 0.975 \\ 
0.60 & 0.114 & 0.220 & 0.145 & 0.276 & 0.437 & 0.479 \\ 
0.70 & 0.072 & 0.095 & 0.074 & 0.122 & 0.086 & 0.103 \\ 
0.90 & 0.052 & 0.074 & 0.060 & 0.065 & 0.047 & 0.065 \\ 
0.95 & 0.070 & 0.122 & 0.061 & 0.094 & 0.065 & 0.102 \\ 
0.99 & 0.096 & 0.200 & 0.076 & 0.113 & 0.097 & 0.192 \\ \hline\hline
\end{tabular}
\end{table}%

\begin{table}[hbp] \centering%
\caption{Empirical power of Quasi LM test based on RE Likelihood; $H_{0}:\rho=0.8$; Nominal size is 0.05; T=9; 2500
replications.\label{key}}%
\begin{tabular}{||l|c|c|c|c|c|c||}
\hline\hline
model & \multicolumn{2}{|c|}{S-Normal} & \multicolumn{2}{|c|}{S-$\chi ^{2}$}
& \multicolumn{2}{|c||}{NS-Normal} \\ \hline
true $\rho $ & $N=100$ & $N=250$ & $N=100$ & $N=250$ & $N=100$ & $N=250$ \\ 
\hline
0.50 & 0.998 & 1.000 & 0.995 & 1.000 & 0.996 & 1.000 \\ 
0.60 & 0.944 & 1.000 & 0.943 & 1.000 & 0.914 & 0.999 \\ 
0.70 & 0.456 & 0.810 & 0.523 & 0.874 & 0.419 & 0.747 \\ 
0.90 & 0.215 & 0.553 & 0.299 & 0.658 & 0.198 & 0.488 \\ 
0.95 & 0.456 & 0.888 & 0.476 & 0.832 & 0.426 & 0.864 \\ 
0.99 & 0.730 & 0.992 & 0.561 & 0.907 & 0.730 & 0.992 \\ \hline\hline
\end{tabular}
\end{table}%

\begin{table}[hbp] \centering%
\caption{Empirical power of Quasi LM test based on FE Likelihood; $H_{0}:\rho=0.8$; Nominal size is 0.05; T=9; 2500
replications.\label{key}}%
\begin{tabular}{||l|c|c|c|c|c|c||}
\hline\hline
model & \multicolumn{2}{|c|}{S-Normal} & \multicolumn{2}{|c|}{S-$\chi ^{2}$}
& \multicolumn{2}{|c||}{NS-Normal} \\ \hline
true $\rho $ & $N=100$ & $N=250$ & $N=100$ & $N=250$ & $N=100$ & $N=250$ \\ 
\hline
0.50 & 0.999 & 1.000 & 0.992 & 1.000 & 0.996 & 1.000 \\ 
0.60 & 0.939 & 0.999 & 0.939 & 0.999 & 0.915 & 0.999 \\ 
0.70 & 0.426 & 0.772 & 0.475 & 0.826 & 0.425 & 0.749 \\ 
0.90 & 0.210 & 0.541 & 0.257 & 0.569 & 0.198 & 0.488 \\ 
0.95 & 0.453 & 0.887 & 0.438 & 0.792 & 0.427 & 0.864 \\ 
0.99 & 0.730 & 0.992 & 0.553 & 0.899 & 0.731 & 0.992 \\ \hline\hline
\end{tabular}
\end{table}%

\begin{table}[hbp] \centering%
\caption{Empirical size of Quasi LM test based on RE Likelihood; Nominal size is 0.05; T=4; $\sigma _{\mu }^{2}=25;$ 2500
replications.\label{key}}%
\begin{tabular}{||l|c|c|c|c|c|c||}
\hline\hline
model & \multicolumn{2}{|c|}{S-Normal} & \multicolumn{2}{|c|}{S-$\chi ^{2}$}
& \multicolumn{2}{|c||}{NS-Normal} \\ \hline
$\rho $ & $N=100$ & $N=250$ & $N=100$ & $N=250$ & $N=100$ & $N=250$ \\ \hline
0.20 & .0472 & .0580 & .0472 & .0504 & .0500 & .0512 \\ 
0.50 & .0524 & .0504 & .0460 & .0468 & .0436 & .0500 \\ 
0.80 & .0564 & .0492 & .0596 & .0564 & .0540 & .0428 \\ 
0.90 & .0528 & .0492 & .0416 & .0496 & .0524 & .0520 \\ 
0.95 & .0540 & .0508 & .0448 & .0504 & .0472 & .0556 \\ 
0.98 & .0460 & .0520 & .0440 & .0448 & .0588 & .0476 \\ 
0.99 & .0460 & .0524 & .0476 & .0504 & .0472 & .0484 \\ \hline\hline
\end{tabular}
\end{table}%

\begin{table}[hbp] \centering%
\caption{Empirical size of Quasi LM test based on RE Likelihood; Nominal size is 0.05; T=9; $\sigma _{\mu }^{2}=25;$ 2500
replications.\label{key}}%
\begin{tabular}{||l|c|c|c|c|c|c||}
\hline\hline
model & \multicolumn{2}{|c|}{S-Normal} & \multicolumn{2}{|c|}{S-$\chi ^{2}$}
& \multicolumn{2}{|c||}{NS-Normal} \\ \hline
$\rho $ & $N=100$ & $N=250$ & $N=100$ & $N=250$ & $N=100$ & $N=250$ \\ \hline
0.20 & .0452 & .0508 & .0480 & .0476 & .0524 & .0472 \\ 
0.50 & .0596 & .0508 & .0512 & .0504 & .0548 & .0488 \\ 
0.80 & .0576 & .0564 & .0564 & .0524 & .0588 & .0584 \\ 
0.90 & .0480 & .0644 & .0576 & .0512 & .0552 & .0504 \\ 
0.95 & .0540 & .0512 & .0624 & .0508 & .0552 & .0548 \\ 
0.98 & .0504 & .0512 & .0504 & .0580 & .0584 & .0664 \\ 
0.99 & .0540 & .0544 & .0532 & .0480 & .0632 & .0456 \\ \hline\hline
\end{tabular}
\end{table}%

\begin{table}[hbp] \centering%
\caption{Empirical power of Quasi LM test based on RE Likelihood; $H_{0}:\rho=0.8$; Nominal size is 0.05; T=4; $\sigma _{\mu }^{2}=25;$ 2500
replications.\label{key}}%
\begin{tabular}{||l|c|c|c|c|c|c||}
\hline\hline
model & \multicolumn{2}{|c|}{S-Normal} & \multicolumn{2}{|c|}{S-$\chi ^{2}$}
& \multicolumn{2}{|c||}{NS-Normal} \\ \hline
$\rho $ & $N=100$ & $N=250$ & $N=100$ & $N=250$ & $N=100$ & $N=250$ \\ \hline
0.50 & .205 & .348 & .238 & .408 & .883 & .982 \\ 
0.60 & .129 & .238 & .136 & .300 & .462 & .501 \\ 
0.70 & .078 & .110 & .076 & .128 & .098 & .102 \\ 
0.90 & .052 & .078 & .055 & .066 & .048 & .075 \\ 
0.95 & .066 & .115 & .060 & .098 & .059 & .123 \\ 
0.99 & .088 & .197 & .076 & .127 & .093 & .203 \\ \hline\hline
\end{tabular}
\end{table}%

\begin{table}[hbp] \centering%
\caption{Empirical power of Quasi LM test based on RE Likelihood; $H_{0}:\rho=0.8$; Nominal size is 0.05; T=9; $\sigma _{\mu }^{2}=25;$ 2500
replications.\label{key}}%
\begin{tabular}{||l|c|c|c|c|c|c||}
\hline\hline
model & \multicolumn{2}{|c|}{S-Normal} & \multicolumn{2}{|c|}{S-$\chi ^{2}$}
& \multicolumn{2}{|c||}{NS-Normal} \\ \hline
$\rho $ & $N=100$ & $N=250$ & $N=100$ & $N=250$ & $N=100$ & $N=250$ \\ \hline
0.50 & .999 & 1.000 & .993 & 1.000 & .997 & 1.000 \\ 
0.60 & .919 & 1.000 & .930 & .999 & .914 & 1.000 \\ 
0.70 & .430 & .784 & .502 & .818 & .398 & .756 \\ 
0.90 & .211 & .544 & .260 & .582 & .188 & .502 \\ 
0.95 & .456 & .872 & .452 & .813 & .438 & .860 \\ 
0.99 & .731 & .992 & .570 & .898 & .730 & .992 \\ \hline\hline
\end{tabular}
\end{table}%

\end{document}